\documentclass[prd,twocolumn,showpacs,letterpaper,nofootinbib,superscriptaddress]{revtex4}
\usepackage{graphics}
\usepackage{epsfig} 
\usepackage{amsmath}
\usepackage{multirow}
\usepackage{rotating}
\usepackage{hyperref}
\usepackage{float}

\input epsf

\newcommand{\be}{\begin{equation}}
\newcommand{\ee}{\end{equation}}
\newcommand{\bea}{\begin{eqnarray}}
\newcommand{\eea}{\end{eqnarray}}
\def\pslash{{\cal P}{\hbox{\kern-6pt $\slash$}}}

\long\def\comment#1{}
\begin{document}
\title{Power law cosmology model comparison with CMB scale information}
\author{Isaac Tutusaus}
\email{isaac.tutusaus@irap.omp.eu}
\affiliation{Universit\'e de Toulouse, UPS-OMP, IRAP, F-31400 Toulouse,  France}
\affiliation{CNRS, IRAP, 14, avenue Edouard Belin, F-31400 Toulouse, France}
\author{Brahim Lamine}
\affiliation{Universit\'e de Toulouse, UPS-OMP, IRAP, F-31400 Toulouse,  France}
\affiliation{CNRS, IRAP, 14, avenue Edouard Belin, F-31400 Toulouse, France}
\author{Alain Blanchard}
\affiliation{Universit\'e de Toulouse, UPS-OMP, IRAP, F-31400 Toulouse,  France}
\affiliation{CNRS, IRAP, 14, avenue Edouard Belin, F-31400 Toulouse, France}
\author{Arnaud Dupays}
\affiliation{Universit\'e de Toulouse, UPS-OMP, IRAP, F-31400 Toulouse,  France}
\affiliation{CNRS, IRAP, 14, avenue Edouard Belin, F-31400 Toulouse, France}
\author{Yves Zolnierowski}
\affiliation{Laboratoire d'Annecy-le-Vieux de Physique des Particules,
CNRS/IN2P3 and Universit\'e Savoie Mont Blanc,
9 Chemin de Bellevue, BP 110, F-74941 Annecy-le-Vieux cedex, France}

\author{Johann Cohen-Tanugi}
\affiliation{Laboratoire Univers et Particules de Montpellier,
  Universit\'{e} de Montpellier, CNRS/IN2P3 Montpellier, 34095
  Montpellier cedex 05, France}
\affiliation{Laboratoire de Physique Corpusculaire, Universit\'{e}
  Clermont Auvergne, Universit\'{e} Blaise Pascal, CNRS/IN2P3
  Clermont-Ferrand, 63178 Aubi\`{e}re cedex, France}

\author{Anne Ealet}
\affiliation{Aix Marseille Univ., CNRS, CPPM, Marseille, 13288
  Marseille cedex 09, France}

\author{St\'{e}phanie Escoffier}
\affiliation{Aix Marseille Univ., CNRS, CPPM, Marseille, 13288
  Marseille cedex 09, France}

\author{Olivier Le F\`{e}vre}
\affiliation{Aix Marseille Univ., CNRS, LAM, UMR 7326, 13388 Marseille, France}

\author{St\'{e}phane Ili\'{c}}
\affiliation{Universit\'e de Toulouse, UPS-OMP, IRAP, F-31400 Toulouse,  France}
\affiliation{CNRS, IRAP, 14, avenue Edouard Belin, F-31400 Toulouse, France}
\affiliation{Aix Marseille Univ., Universit\'e de Toulon, CNRS, CPT,
  13288 Marseille cedex 09, France}

\author{Alice Pisani}
\affiliation{Aix Marseille Univ., CNRS, CPPM, Marseille, 13288
  Marseille cedex 09, France}
\affiliation{Sorbonne Universit\'{e}s, UPMC Univ. Paris 06, UMR 7095,
Institut d’Astrophysique de Paris, 98 bis boulevard Arago, F-75014
Paris, France}
\affiliation{CNRS, UMR 7095, Institut d’Astrophysique de Paris, 98 bis
  boulevard Arago, F-75014 Paris, France}

\author{St\'{e}phane Plaszczynski}
\affiliation{Laboratoire de l'Acc\'el\'erateur Lin\'eaire,
  Univ. Paris-Sud, CNRS/IN2P3, Universit\'e Paris-Saclay, 91898 Orsay cedex, France}

\author{Ziad Sakr}
\affiliation{Universit\'e de Toulouse, UPS-OMP, IRAP, F-31400 Toulouse,  France}
\affiliation{CNRS, IRAP, 14, avenue Edouard Belin, F-31400 Toulouse, France}
\affiliation{Faculty of Sciences, Universit\'{e} St Joseph, UR EGFEM,
  Beirut 1107 2050, Lebanon}

\author{Valentina Salvatelli}
\affiliation{Aix Marseille Univ., Universit\'e de Toulon, CNRS, CPT,
  13288 Marseille cedex 09, France}

\author{Thomas Sch\"{u}cker}
\affiliation{Aix Marseille Univ., Universit\'e de Toulon, CNRS, CPT,
  13288 Marseille cedex 09, France}

\author{Andr\'{e} Tilquin}
\affiliation{Aix Marseille Univ., CNRS, CPPM, Marseille, 13288
  Marseille cedex 09, France}

\author{Jean-Marc Virey}
\affiliation{Aix Marseille Univ., Universit\'e de Toulon, CNRS, CPT,
  13288 Marseille cedex 09, France}

\date{\today}

\begin{abstract}
 Despite the ability of the cosmological concordance model ($\Lambda$CDM) to describe the cosmological
 observations exceedingly well, power law expansion of the Universe
 scale radius, $R(t) \propto t^n $, has been
 proposed as an alternative framework. We examine here 
 these models, analyzing their ability to fit cosmological data
 using robust model comparison criteria. Type Ia
 supernovae (SNIa), baryonic acoustic oscillations (BAO) and acoustic scale information from the cosmic microwave background (CMB) have
 been used.  We find that SNIa data either alone or combined with BAO can be well reproduced  by  both $\Lambda$CDM and power law
 expansion models with $n \sim 1.5$, while the constant expansion rate
 model ($n = 1$) is clearly disfavored. Allowing
 for some redshift evolution in the  SNIa luminosity essentially
 removes any clear preference for a specific model.
The CMB data are well known to provide the most stringent constraints on standard cosmological models,
in particular, through the position of the first peak of the
 temperature angular power spectrum, corresponding to the sound horizon at recombination, a scale
physically related to the BAO scale. Models with $n \geq 1$ lead to a
divergence of the sound horizon and do not
naturally provide the relevant scales for the BAO and the CMB. We
retain an empirical footing to overcome this issue: we let the data
choose the preferred values for these scales, while we recompute the
ionization history in power law models, to obtain the distance to the
CMB. In doing so, we find that
the scale coming from the BAO data is not consistent with the observed
position of the first peak of the CMB temperature angular power
spectrum for any power law  cosmology. Therefore, we conclude that
when the three standard probes (SNIa, BAO, and CMB) are
combined, the $\Lambda$CDM model is very strongly favored over any of these
alternative models, which are then essentially ruled out.
\end{abstract}

\maketitle
\vskip 1cm

\section{INTRODUCTION}

The cosmological concordance model ($\Lambda$CDM) framework offers a simple description of the
properties of our Universe with a very small number of free
parameters, reproducing remarkably well a wealth of high quality
observations (allowing us to reach a precision below
5\% for most of the parameters with present-day data\,\cite{Planck2015table}).
More than fifteen years after the discovery of the accelerated
expansion of the Universe\,\cite{Riess,Perlmutter}, the $\Lambda$CDM model remains the current standard model in
cosmology. However, since the dark contents of the Universe remain unidentified, alternative models still deserve to be investigated.

A notable alternative to the $\Lambda$CDM model is the
so-called power law cosmology, where the scale factor $a(t)$ evolves
proportionally to a power of the proper time: $a(t)\propto t^n$. This
class of models may for instance emerge when classical fields couple to spacetime
curvature\,\cite{PLtheory}. Predicted abundances by primordial nucleosynthesis
seem problematic \cite{BBN,BBN2}, but the confrontation to 
low-redshift data, type Ia
supernovae (SNIa) and the baryonic acoustic oscillations (BAO) may not be as problematic\,\cite{Dolgov}, although there is some controversy in the
literature concerning the ability of the power law
cosmology to fit these data \cite{Shafer}.
It seems therefore 
 interesting to compare  the performance of the standard
$\Lambda$CDM model to those of power law models  taking into account standard cosmological  probes.

Among these alternative models to $\Lambda$CDM stands the so-called $R_h=ct$
cosmology\,\cite{Melia2012}, where $R_h=c/H(t)$ is the Hubble
radius and $H(t)$ the Hubble parameter. This model, which is characterized by a total
equation of state $\rho+3p=0$, turns out to be a particular case of the power law
cosmology with exponent equal to 1. From a theoretical point of view, there is also 
some controversy on the motivation for such models\,\cite{Meliath1,Meliath2,Meliath3,Lewis2013}.
As in the general power law case,
some studies claimed that this model is ruled out by
observations\,\cite{Bilicki,Shafer}, while some others claimed that $R_h=ct$ is able to fit the data even better
than $\Lambda$CDM\,\cite{MeliaSNIa}, and that it can explain a large
amount of physics like the epoch of reionization\,\cite{Melia1}, the
high-redshift quasars\,\cite{Melia2}, the cosmic microwave background (CMB) multipole
alignment\,\cite{Melia3} or the constancy of the cluster gas mass fraction\,\cite{Melia4}.

In the following, we examine how the $\Lambda$CDM and  power
law models   compare to the main cosmological probes, 
using robust model selection
criteria. In this work, with respect to previous ones,  we allow some evolution with redshift for the  SNIa luminosity (considering
an evolution in the distance modulus as a function of the redshift) and consider the implication of the 
 CMB properties, which certainly represents the most impressive success of the standard model, and use it 
in combination with the above-mentioned low-redshift probes.

In Sec.\,\ref{section2} we briefly describe the models under study. In Sec.\,\ref{section3}
we present the statistical tool used to determine the ability of a model to
fit the data and the selection criteria used for this work. In Sec.\,\ref{section4} we describe the two low-redshift probes used in the study: SNIa and BAO, as well as the data samples used and the parameters that enter
into the comparison. In Sec.\,\ref{CMB} we present the high-redshift probe
used, CMB, and we describe the approach followed in this work in order
to use data coming from this probe. We present the results obtained in
Sec.\,\ref{section6} and conclude in Sec.\,\ref{section7}.

\section{MODELS}\label{section2}

In this section we present the three different models studied in this
work: the $\Lambda$CDM model, the power law cosmology and the $R_h=ct$
cosmology.

\subsection{$\Lambda$CDM model}

The flat $\Lambda$CDM model is the current standard model
in cosmology thanks to its adequacy with the main 
cosmological
data, i.e.   SNIa\,\cite{Betoule2014}, BAO\,\cite{Anderson2014} and CMB\,\cite{Planck2015table}. This model assumes a Robertson-Walker
metric and Friedmann-Lema\^{i}tre dynamics leading to the comoving
angular diameter distance, $r(z)$, and Friedmann-Lema\^{i}tre equation,

\begin{align}
  r(z)&=c\int_0^z\frac{\text{d}z'}{H(z')}\,,\\
\frac{H^2(z)}{H_0^2}&=\Omega_r(1+z)^4+\Omega_m(1+z)^3+(1-
\Omega_r-\Omega_m)\,,\label{eq2}
\end{align}
where $H_0$ is the Hubble constant and $\Omega_i$ is the energy density
parameter of the fluid $i$. The Universe flatness is already captured
by the last term in Eq.\,(\ref{eq2}). We compute the radiation contribution as\,\cite{Planck2015table},

\begin{equation}
\Omega_r=\Omega_{\gamma}\left[1+N_{\text{eff}}\,\frac{7}{8}\left(\frac{4}{11}\right)^{4/3}\right]\,,
\end{equation}
where $\Omega_{\gamma}$, the photon contribution, is given by,

\begin{equation}\label{Oph}
\Omega_{\gamma}=4\cdot 5.6704\times
    10^{-8}\frac{T_{\text{cmb}}^4}{c^3}\frac {8\pi G}{3H_0^2}\,,
\end{equation}
and fixing\footnote{We have checked that small variations on these
  parameters do not affect the results.} $N_{\text{eff}}=3.04$\,\cite{Planck2015table}, the effective number of
neutrinolike relativistic degrees of freedom, $H_0=67.74\,\text{km}\,\text{s}^{-1}\text{Mpc}^{-1}$\,\cite{Planck2015table}
and $T_{\text{cmb}}=2.725$\,\cite{COBE}, the temperature of the CMB today. Notice
that we only fix $H_0$ for the radiation contribution. It is left free
in the rest of the work.

\subsection{Power law and $R_h=ct$ cosmologies}

The main assumption in power law cosmologies is that the scale factor
evolves as a power of the proper time,

\begin{equation}
  a(t)=\left(\frac{t}{t_0}\right)^n\,,
\end{equation}
where $n$ is the power of the model and $a(t_0)=1$. This provides us
with the Friedmann-Lema\^{i}tre equation,

\begin{equation}
H(z)=H_0(1+z)^{1/n}\,,
\end{equation}
which leads to

\begin{equation}
  r(z)=\frac{c}{H_0}\times \left\{
\begin{array}{cc}
\frac{(1+z)^{1-1/n}-1}{1-1/n}\,, & n\neq 1\,,\\
\ln (1+z)\,, & n=1\,.
\end{array}\right.
\end{equation}

Notice that an expanding Universe requires $0 < n <\infty$.

The $R_h=ct$ cosmology states that 
the Hubble radius $R_h=c/H(t)$ is
 proportional to time, so we
can write $H(t)=t^{-1}$ for this model. For  a flat Universe this leads to the comoving
angular diameter distance and Friedmann-Lema\^{i}tre equation,

\begin{align}
r(z)&=\frac{c}{H_0}\ln(1+z)\,,\\
H(z)&=H_0(1+z)\,.
\end{align}

\begin{figure}
\includegraphics[scale=.45]{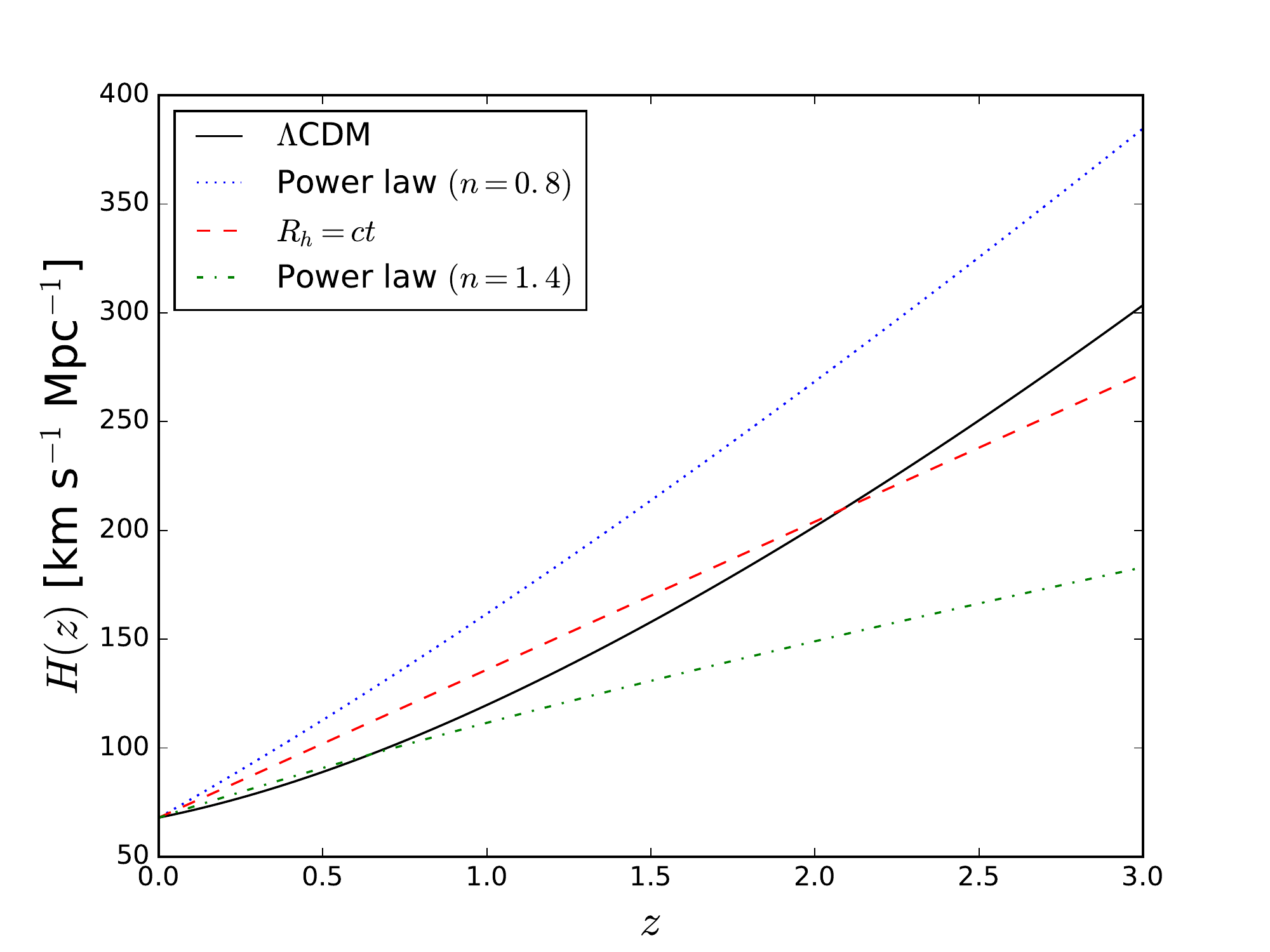}
\caption{Hubble parameter as a function of the redshift for
  $\Lambda$CDM, $R_h=ct$ cosmology and two different power law
  cosmologies. $\Omega_m$, $\Omega_r$ and $H_0$ have been fixed to
  $0.3$, $8\times 10^{-5}$ and
  $68\,\text{km}\,\text{s}^{-1}\text{Mpc}^{-1}$, respectively, for illustrative purposes.}\label{fig000}
\end{figure}

Figure\,\ref{fig000} shows the variation with redshift of the Hubble
parameter $H(z)$ for these three models. The matter and
radiation contributions have been fixed to $0.3$ and $8\times 10^{-5}$,
respectively, while the Hubble constant has been fixed to
$68\,\text{km}\,\text{s}^{-1}\text{Mpc}^{-1}$, for illustrative purposes.

\section{METHOD}\label{section3}
In this section we review the statistical tools used to quantify the goodness
of fit and to compare the models under study.

\subsection{Goodness of fit}
To quantify the capacity of a model to fit the data we minimize the
common $\chi^2$ function,

\begin{equation}\label{eqchi2}
\chi^2=(\textbf{u}-\textbf{u}_{data})^TC^{-1}(\textbf{u}-\textbf{u}_{data})\,,
\end{equation}
 using the MIGRAD application from the \texttt{iminuit} Python package\footnote{\url{https://github.com/iminuit/iminuit}}, designed for finding the minimum value of a multiparameter
function and analyzing the shape of the function around the
minimum. This code is the Python implementation of the
former MINUIT Fortran code\,\cite{minuit}. In Eq.\,(\ref{eqchi2}), $\textbf{u}$
stands for the model prediction, while $\textbf{u}_{data}$ and
$C^{-1}$ hold for the observables and their inverse covariance matrix,
respectively. We
then compute the probability that a larger
value for the $\chi^2$ could occur for a fit with $\nu=N-k$ degrees
of freedom, where $N$ is the number of data points and $k$ is the
number of free parameters of the model,

\begin{equation}\label{Prob}
P(\chi^2,\nu)=\frac{\Gamma\left(\frac{\nu}{2},\frac{\chi^2}{2}\right)}
{\Gamma\left(\frac{\nu}{2}\right)}\,,
\end{equation}
with $\Gamma(t,x)$ being the upper incomplete gamma function and
$\Gamma(t)=\Gamma(t,0)$ the complete gamma function.

Obtaining a probability close to 1 implies that it is very
likely to get larger $\chi^2$ values, meaning that the model fits
correctly (possibly too well) the data. On the other hand, obtaining a small probability indicates
that the model does not provide a good fit to the data.

When combining probes, we minimize the $\chi^2$ function given by the
sum of individual $\chi^2$ functions, i.e., we assume that the
probes are statistically independent.

\begin{figure*}
\includegraphics[scale=0.45]{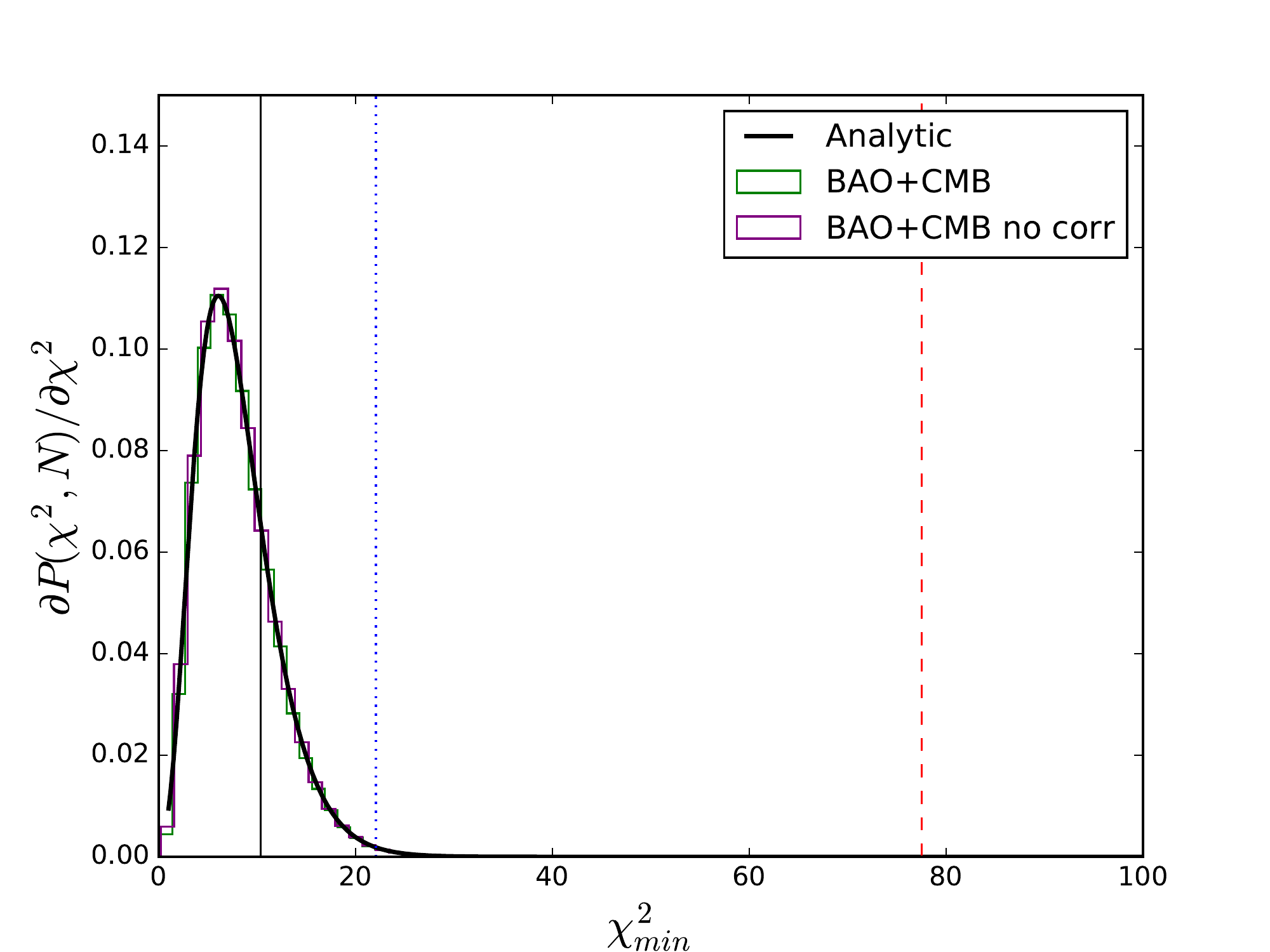}\,\includegraphics[scale=0.45]{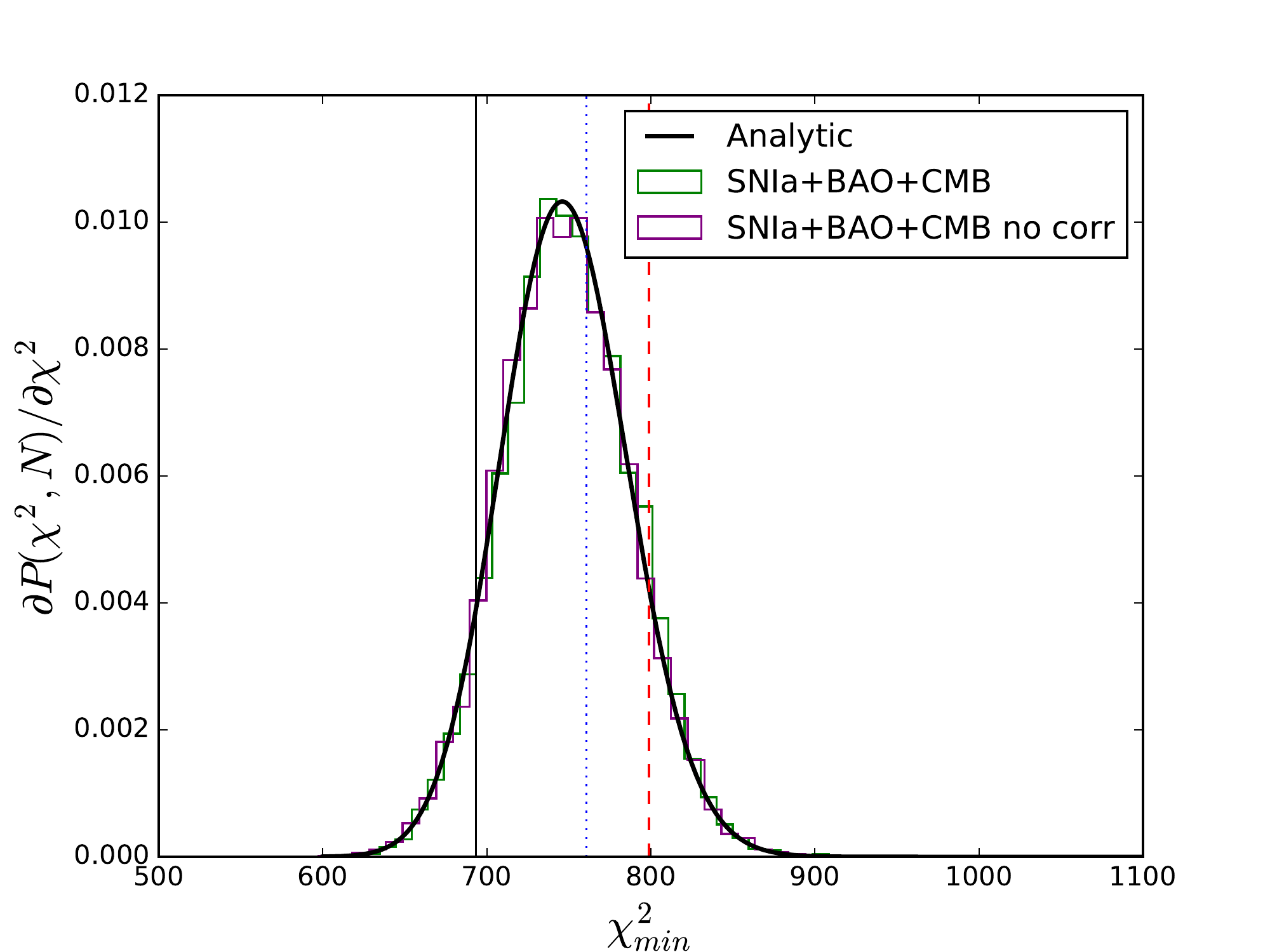}
\caption{Histograms of $\chi^2$ for Monte Carlo simulations [to study
  the impact of correlations in Eq.\,(\ref{Prob})] using correlations
  (green) and neglecting them (purple). The analytic distribution is
  also represented for further comparison (thick black solid
  line). The compatibility of the three distributions in each plot
  shows that Eq.\,(\ref{Prob}) can be used in this work. The measured
  values of the minimum
  of the $\chi^2$ are also represented, only for illustrative
  purposes, for each model and each combination of probes used (see
  Table\,\ref{table2}. Black solid
  line, $\Lambda$CDM; blue dotted line, power law cosmology; red
  dashed line, $R_h=ct$ cosmology). Left plot: BAO+CMB covariance matrix with
  $M$=100000 iterations. Right plot: SNIa+BAO+CMB covariance matrix
  with $M$=10000 iterations.}\label{fig4bis}
\end{figure*}

It is important to notice that Eq.\,(\ref{Prob}) is only valid when we
work with $N$ data points coming from $N$ independent random variables
with Gaussian distributions. However, in this work we consider the
correlation within probes; thus, the data points come from nonindependent Gaussian random variables. In order to check the impact of
correlations on this probability we compute the histogram of $\chi^2$ through Monte Carlo
simulations with and without correlations. First of all, we fix the fiducial model to
$\textbf{u}=\textbf{0}$ in order to save computation time, since we do
not need then to fit a
certain model each time we compute a $\chi^2$. Notice that there are
no parameters then; thus, $k=0$ and $\nu=N$. We then generate the
data set from an $N$-dimensional Gaussian distribution centered at $\textbf{0}$
and with the corresponding
covariance matrix $C$ for the probes used. When neglecting
correlations we consider only the diagonal terms of $C$. Finally, we compute the $\chi^2$
using Eq.\,(\ref{eqchi2}) and we repeat $M$ times to obtain
the histograms shown in Fig.\,\ref{fig4bis}.

In the left plot we
use the covariance matrix for BAO and CMB data. We can clearly observe
that the histogram obtained with correlations (green) is completely consistent
with the histogram obtained neglecting any correlation (purple). Moreover, both
of them are consistent with the analytic distribution (thick black solid line), which is given
by the derivative of Eq.\,(\ref{Prob}) with respect to
$\chi^2$. Notice that we have neglected here the number of free parameters
of the model because we have fixed the fiducial model to
$\textbf{0}$. The fact that the three distributions are completely
consistent implies that the correlations in the BAO+CMB covariance
matrix do not effect Eq.\,(\ref{Prob}) and we can safely use it. In
the right plot of Fig.\,\ref{fig4bis} we have the equivalent results
using the covariance matrix for SNIa, BAO and CMB. As before, the three
distributions are completely consistent, implying that we can use
Eq.\,(\ref{Prob}) with these correlations. A particularity in this
case is that the covariance matrix is not completely independent of
the cosmology. As is discussed in the following sections, the
covariance matrix depends on two nuisance parameters. In order to correctly predict the
effect of these correlations, we need to consider these nuisance
parameters and determine them when fitting each model under study to
the $M$ mock data samples. However, we keep the fiducial
$\textbf{u}=\textbf{0}$ model, due to the fact that the two SNIa nuisance parameters remain very close to $\alpha=0.14$
and $\beta=3.1$, as can be seen in Table\,\ref{table1}.

\subsection{Model comparison}\label{modelcomparison}
In this work we consider two
widely used criteria to compare the models under study: the Akaike
information criterion (AIC)\,\cite{AIC} and the Bayesian information
criterion (BIC)\,\cite{BIC}. Both account for the fact that a
model with fewer parameters is generally preferable to a more complex model if
both of them  fit the data equally well.

The AIC is built from information theory. Rather than having a simple
measure of the direct distance between two models (Kullback-Leibler
distance\footnote{The Kullback-Leibler information between models $f$
  and $g$ denotes the information lost when $g$ is used to approximate
$f$. As a heuristic interpretation, the Kullback-Leibler information
is the distance from $g$ to $f$.}), the AIC provides us with an estimate of the expected, relative
distance between the fitted model and the unknown true mechanism that
actually generated the observed data. We must be aware that the AIC is
useful in selecting the best model in the set of tested models;
however, if all the models are very poor, the AIC still gives us
the one estimated to be the best. This is why we previously computed
the probability of a model to correctly fit the data [see
Eq.\,(\ref{Prob})]. Given the minimum of the $\chi^2$ ($\chi^2_{min}$) and the
number of free parameters of the model $k$, the AIC is given by

\begin{equation}
\text{AIC}=\chi^2_{min}+2k\,.
\end{equation}

The AIC may perform poorly if there are too many parameters compared
to the size of the sample\,\cite{AICc1,AICc2}. In this case, a second-order
variant of the AIC can be used, the so-called AICc\,\cite{AICc25},

\begin{equation}\label{aicc}
\text{AICc}=\text{AIC}+\frac{2k(k+1)}{N-k-1}\,,
\end{equation}
where $N$ is the number of data points. An extensive presentation and
discussion of the AIC and its variations can be found in\,\cite{AICc3}.

The BIC is one of the most used criterion from the so-called
dimension-consistent criteria (see \cite{BIC1} for a review of many of
these criteria). It was derived in a Bayesian context with equal prior
probability on each model and minimal priors on the parameters,
given the model. It is given by

\begin{equation}\label{bic}
\text{BIC}=\chi^2_{min}+k\ln (N)\,.
\end{equation}

Both the AICc and the BIC strongly depend on
the size of the sample. In order to compare different models we use the exponential of the differences
$\Delta$AICc/2 ($\Delta$BIC/2), where
$\Delta$AICc=$\text{AICc}_{\Lambda\text{CDM}}$-AICc (Id. for the BIC),
since the
exponential can be interpreted as the relative probability that the
corresponding model minimizes the estimated information loss with
respect to the $\Lambda$CDM model.

Given that the AIC (AICc) and the BIC can both be derived as either frequentist
or Bayesian procedures, what fundamentally distinguishes them is their
different philosophy, including the nature of their target
models. Thus, the choice of the criterion depends on their performance under
realistic conditions. A comparison of these two criteria is outside
the scope of this paper (see\,\cite{COMP} for an extended and
detailed comparison), so we just provide the results for both of them. In
general, though, the BIC penalizes extra parameters more severely than the
AIC.

It is important to notice that when comparing two models with the same
data sample and the same number of parameters (e.g. $\Lambda$CDM and power
law with SNIa data, $N=740$, $k=5$), $\Delta$AIC and $\Delta$BIC basically
reduce to $\Delta\chi^2_{min}=\chi^2_{\Lambda\text{CDM}}-\chi^2$. This
leads to the same numerical values for exp$(\Delta \text{AICc}/2)$ and
exp$(\Delta \text{BIC}/2)$ for $\Lambda$CDM and power law models in Table\,\ref{table2}.

\section{Low-redshift probes}\label{section4}

In this section we describe the two low-redshift cosmological
probes and the corresponding data
sets used to compare the different models presented.

\subsection{SNIa}

Type Ia supernovae are considered as standardizable candles useful
to measure cosmological distances. Although measurements of CMB and
large-scale structure can constrain the matter content of the Universe
and the dark energy equation of state parameter, SNIa are important
for breaking degeneracies and achieve precise cosmological
measurements. The observable used in SNIa measurements is the
so-called distance modulus,

\begin{equation}
  \mu(z)=5\log_{10}\left(\frac{H_0}{c}d_L(z)\right)\,,
\end{equation}
where $d_L(z)=(1+z)r(z)$ is the luminosity distance. Notice that we
have defined the
distance modulus in such a way that it is independent of the $H_0$ parameter, which is degenerate with the
SNIa absolute magnitude.

Distance estimation with SNIa is based on empirical observation that
these events form a homogeneous class whose variability can be
characterized by two parameters\,\cite{Tripp1998}: the time stretching
of the light curve ($X_1$) and the supernova color at maximum
brightness ($C$). In this work we use the joint light-curve analysis
for SNIa from \cite{Betoule2014}. The authors assume that supernovae
with identical color, shape and galactic environment have on average
the same intrinsic luminosity for all redshifts. This yields the
distance modulus,
\begin{equation}
\mu_{\text{obs}}=m_B^*-(M_B-\alpha\times X_1+\beta \times C)\,,
\end{equation}
where $m_B^*$ corresponds to the observed peak magnitude in the
rest-frame B band and $\alpha$ and $\beta$ are nuisance parameters
in the distance estimate. The $M_B$ nuisance parameter is given by the
step function,
\begin{equation}
M_B=\left\{
\begin{array}{cl}
M_B^1\,, & \text{if } M_{\text{stellar}} < 10^{10} M_{\odot}\,,\\
M_B^1+\Delta M\,, & \text{otherwise}\,,
\end{array}\right.
\end{equation}
where $M_B^1$ and $\Delta M$ are nuisance parameters, in order to take
into account the dependence on host galaxy properties.

Concerning the errors and correlations on the measurements we use the
covariance
matrix\footnote{\url{http://supernovae.in2p3.fr/sdss_snls_jla/}}
provided by \cite{Betoule2014} where the authors consider
the contribution from error propagation of light-curve
fit uncertainties (statistical contribution) and the contribution of
seven sources of systematic uncertainty: the calibration, the light-curve
model, the bias correction, the mass step, the dust extinction, the
peculiar velocities and the contamination of nontype Ia supernovae.

In some specific cases during this work we relax the redshift
independence assumption made in \cite{Betoule2014}. In order to
account for a possible SNIa evolution with redshift (caused by
some astrophysical procedures, for
example, see\,\cite{SNIaevI,SNIaevII} for previous studies accounting
for SNIa evolution) we add an extra nuisance
parameter $\epsilon$ to the distance modulus estimate,
\begin{equation}\label{ev}
\mu_{\text{obs}}=m_B^*-(M_B-\alpha\times X_1+\beta \times
C-\epsilon\times z)\,.
\end{equation}

When using SNIa data, the set of nuisance parameters considered is
$\{\alpha,\,\beta,\,M,\,\Delta M,\,\epsilon\}$. For $\Lambda$CDM and the
power law cosmology we
consider $\Omega_m$ and $n$, respectively, as cosmological
parameters. We consider no cosmological parameters when using
SNIa data with the $R_h=ct$ cosmology.

\subsection{BAO}

The baryonic acoustic oscillations are the regular and periodic
fluctuations of visible matter density in large-scale structure. They
are characterized by the length of a standard ruler, generally denoted by $r_d$. The
main observable used in BAO measurements is the ratio of the BAO
distance at low redshift to this scale $r_d$. In
the $\Lambda$CDM model, the BAO come from the sound waves propagating
in the early Universe and the standard ruler $r_d$ is equal to the comoving sound horizon at
the redshift of the baryon drag epoch: $r_d=r_s(z_d)$, $z_d\approx 1060$. For models
differing from the $\Lambda$CDM model, $r_d$ does not need to coincide with
$r_s(z_d)$\,\cite{Verde}. For the moment, and in order to be as general as possible, we do not
delve into the physics governing the sound horizon $r_d$, so we
consider $r_d$ as a free parameter.

The BAO are usually
assumed to be isotropic. In this case the BAO distance scale is given
by
\begin{equation}
D_V(z)\equiv\left(r^2(z)\frac{cz}{H(z)}\right)^{1/3}\,.
\end{equation}

More recently it has also been possible to measure radial and
transverse clustering separately, allowing for anisotropic BAO. The
BAO distance scales are then $r(z)$ and $c/H(z)$.

In this work we follow\,\cite{Aubourg} in combining the measurements of 6dFGS\,\cite{BAO1},
SDSS [Main Galaxy Sample (MGS)]\,\cite{BAO2}, BOSS (CMASS and LOWZ
samples - Data Release 11)\,\cite{BAO3,Anderson2014} and BOSS Lyman-$\alpha$
forest (Data Release 11)\,\cite{BAO5,BAO6}. As in \cite{Shafer}, we assume that all the
measurements are independent, apart from the CMASS anisotropic
measurements (correlated with coefficient -0.52) and the
Lyman-$\alpha$ forest measurements (correlated with coefficient
-0.48).

According to \cite{Bassett}, when
constraining parameters to a high confidence level or claiming that a
model is a poor fit to the data, one should take into account that BAO
observable likelihoods are not Gaussian far from the peak. In this work we follow the same
approach and account for this effect by replacing the usual $\Delta
\chi^2_G=-2\ln \mathcal{L}_G$ for a Gaussian likelihood observable by
\begin{equation}
\Delta\chi^2=\frac{\Delta\chi^2_G}{\sqrt{1+\Delta\chi^4_G\left(\frac{S}{N}\right)^{-4}}}\,,
\end{equation}
where $S/N$ is the corresponding detection significance, in units of
$\sigma$, of the BAO feature. We consider a detection significance of
2.4$\sigma$ for 6dFGS, 2$\sigma$ for SDSS MGS, 4$\sigma$ for BOSS
LOWZ, 6$\sigma$ for BOSS CMASS and 4$\sigma$ for BOSS Lyman-$\alpha$
forest. 

When using BAO data, we consider the following set of cosmological
parameters: $\{r_d\times H_0/c,\,\Omega_m,\,n\}$. The latter two only
apply for $\Lambda$CDM and power law cosmology, respectively. We do
not consider any nuisance parameter.

\section{High-redshift probe: CMB}\label{CMB}
In this section we present the high-redshift probe used, 
the cosmic microwave background, and the approach we follow in
order to consider this probe in our study, including power law models.

The CMB is an extremely powerful source of information due to 
the high  precision of modern data. Furthermore it represents   high
redshift data, complementing low-redshift probes.  In standard cosmology, 
the physics governing the sound horizon at the early
Universe is 
that of a baryon-photon plasma in an expanding Universe. 
The comoving sound horizon at the last scattering redshift is given by
\begin{equation}\label{rs}
r_s(z_{*})=\int_{z_{*}}^{\infty}\frac{c_s(z)\,\text{d}z}{H(z)}\,,
\end{equation}
where $z_*$ stands for the redshift of the last scattering and where
\begin{equation}
c_s(z)=\frac{c}{\sqrt{3(1+R_b(z))}},\hspace{15pt} R_b(z)=\frac{3\rho_b}{4\rho_{\gamma}}\,,
\end{equation}
with $\rho_b$ being the baryon density and $\rho_\gamma$ the photon
density.
 The observed angular scale of the sound horizon at recombination,
\begin{equation}\label{l_a}
\ell_a\equiv \frac{\pi c}{r_s(z_{*})} \int_0^{z_{*}}\frac{\text{d}z}{H(z)} \,,
\end{equation}
then depends on the angular distance to the CMB, a physical quantity sensitive to the expansion history up to $z_{*}$
and thereby to the background history of models\,\cite{CMB1}.  Notice that $\ell_a$
roughly corresponds to the position of the first peak of the
temperature angular power spectrum of the CMB. Although this represents a reduced fraction of the information,
it is well known that reduced parameters capture a large fraction of
the information contained in the CMB fluctuations of the angular power spectra\,\cite{Wang}. 
We
use the value provided by the Planck
Collaboration\,\cite{Planck2015noMG}: $\ell_a=301.63\pm 0.15$ and, in the
following, we refer to this information as CMB data.
 It has
been obtained from Planck temperature and low-$\ell$ polarization data. Marginalization
over the amplitude of the lensing power spectrum has been performed,
since it leads to a more conservative approach. 

According to \cite{Melia2015}, the $R_h=ct$ universe assumes the
presence of dark energy and radiation in addition to baryonic and dark matter. The only requirement of this model is to constrain
the total equation of state by requiring $\rho+3p=0$. Following this
definition, and extending the idea to the power law cosmology, we infer that the physics governing the sound horizon at the early
Universe is the same as for $\Lambda$CDM, since we are again essentially dealing with  a
baryon-photon plasma in an expanding universe.

For $\Lambda$CDM, we use the value provided in\,\cite{Planck2015table} for
$\Omega_b h^2=0.02230$ and we use Eq.\,(\ref{Oph}) for the radiation
contribution. This assumption has already been made in the
literature. In\,\cite{Levy}, for example, the authors considered the Dirac-Milne universe (a matter-antimatter symmetric
cosmology) and kept the same expression for $r_s(z_{*})$ as in
the $\Lambda$CDM case. 

For a power law cosmology,

\begin{equation}
r_s(z_{*})\propto \int_{z_{*}}^{\infty}\frac{(1+z)^{1/2}}{(1+z)^{1/n}}\,\text{d}z\,
\end{equation}
which converges only for $n<2/3$; therefore, there is already a
fundamental problem in these theories when describing the early
Universe. This divergence also exists for the sound horizon
$r_d$ in the BAO. Given that the big bang nucleosynthesis already suffers from a problem in the early Universe,
one might imagine that the physics of the early Universe allows us to solve this issue, essentially by restoring the standard model in the very early Universe, keeping the sound horizon finite. 
$r_d$ being now an unknown quantity, we have to obtain its value by fitting it to the data. 
We can then develop $r_s(z_{*})$ by,

\begin{equation}
r_s(z_{*})=\int_{z_{*}}^{\infty}\frac{c_s(z)\,\text{d}z}{H(z)}=r_d-\int_{z_{d}}^{z_{*}}\frac{c_s(z)\,\text{d}z}{H(z)}\,.
\end{equation}

In\,\cite{Levy} the authors also had to deal with this divergence near the initial
singularity. They opted for putting upper and lower bounds to the
integral on physically motivated grounds, while we allow the data to determine $r_d$ and avoid the divergence.

We now need to determine $z_d$
and $z_*$ for all the models. A common definition of the redshift of
the CMB is given by the maximum of the visibility
function\,\cite{Hu96},

\begin{equation}\label{visibility}
g(z)=\dot{\tau}(z)e^{-\tau(z)}\,,
\end{equation}
where $\tau(z)$ is the optical depth\,\cite{tau},

\begin{equation}
\tau(z)=\sigma_T\int_0^z n_e(z')\frac{c}{(1+z')H(z')}\,\text{d}z'\,,
\end{equation}
with $\sigma_T$ being the Thomson cross section and $n_e$ the free
electron number density. This definition is well motivated because the
visibility function can be understood as the probability of the last
photons of the CMB to scatter; thus, the maximum provides us with the
most probable redshift of this last scattering. In order to obtain
$n_e$ we calculate the free electron fraction $X_e$ and we further multiply it by the hydrogen number density,

\begin{equation}
n_e(z)=X_e(z)\left[\frac{3H_0^2\Omega_b}{8\pi G m_H\mu}(1+z)^3\right]\,,
\end{equation}
where $m_H$ is the mass of the hydrogen atom and $\mu=1/(1-Y)$ with
$Y$ being the helium mass fraction.

\begin{figure}
\includegraphics[scale=.45]{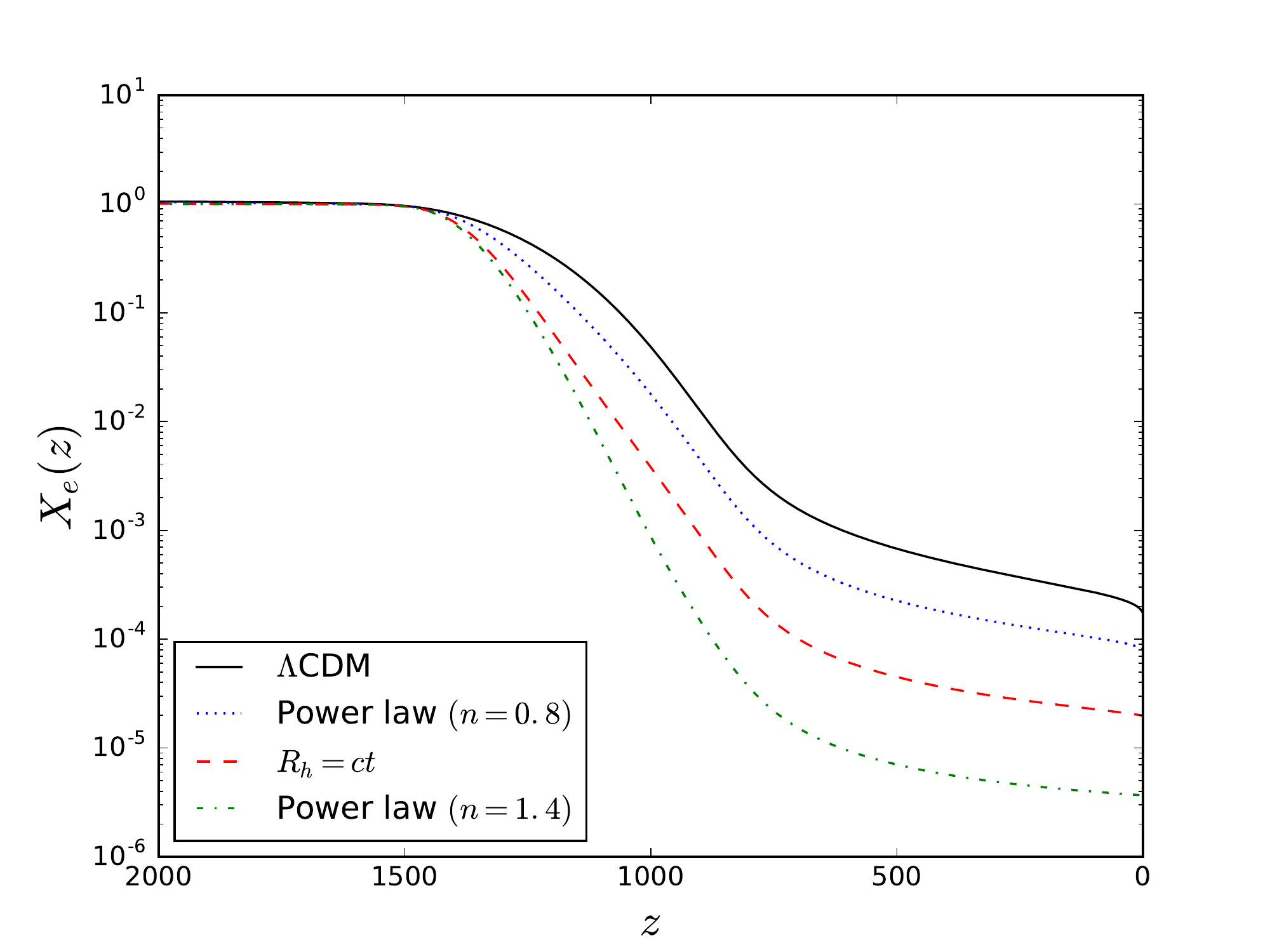}
\caption{Free electron function $X_e$ as a function of the redshift
  for $\Lambda$CDM, $R_h=ct$ cosmology and two different power law
  cosmologies. The parameters relevant for reionization have been
  fixed to the Planck 2015 values for illustrative
  purposes\,\cite{Planck2015table} (helium mass fraction, CMB
  temperature at $z=0$, $\Omega_m$, $\Omega_b$, $\Omega_k$, $h$ and $N_{\text{eff}}$).}\label{figXe}
\end{figure}

The ionization history $X_e(z)$ depends on the expansion
rate. In order to obtain it we use the 
\texttt{Recfast++}\,\cite{Recfast++}
code,\footnote{\url{http://www.cita.utoronto.ca/~jchluba/Science_Jens/Recombination/Recfast++.html}} based on the C version of
\texttt{Recfast}\,\cite{Recfast}, adapting the expansion history to
the corresponding one for each model. This new version includes
recombination corrections\,\cite{Recfast++,RubinoMartin} and allows us to run a dark matter annihilation
module\,\cite{Chluba}. It also includes a new ordinary differential
equations solver\,\cite{ODE}. More details about this code can be
found in\,\cite{Recfast++1,Recfast++2,Recfast++3}. Figure\,\ref{figXe} provides a comparison between $X_e$ for the different
power law cosmologies and $\Lambda$CDM.
We have neglected the recombination corrections and dark
matter annihilations for simplicity, and because this level of
precision in the $X_e$ determination is not needed for our purposes.

Another definition for the redshift of the CMB is the one adopted by
the Planck Collaboration\,\cite{Planck2015table} by determining the
redshift when the optical depth equals 1. We denote $z_{CMB}$ the
redshift obtained with the first definition [Eq.\,(\ref{visibility})]
and $z_*$ the redshift obtained with the Planck Collaboration
convention. Although we use $z_*$ for consistency with Planck
when performing our analyses, we have defined $z_{CMB}$ for
illustrative and comparative purposes.  

\begin{figure}
\includegraphics[scale=.45]{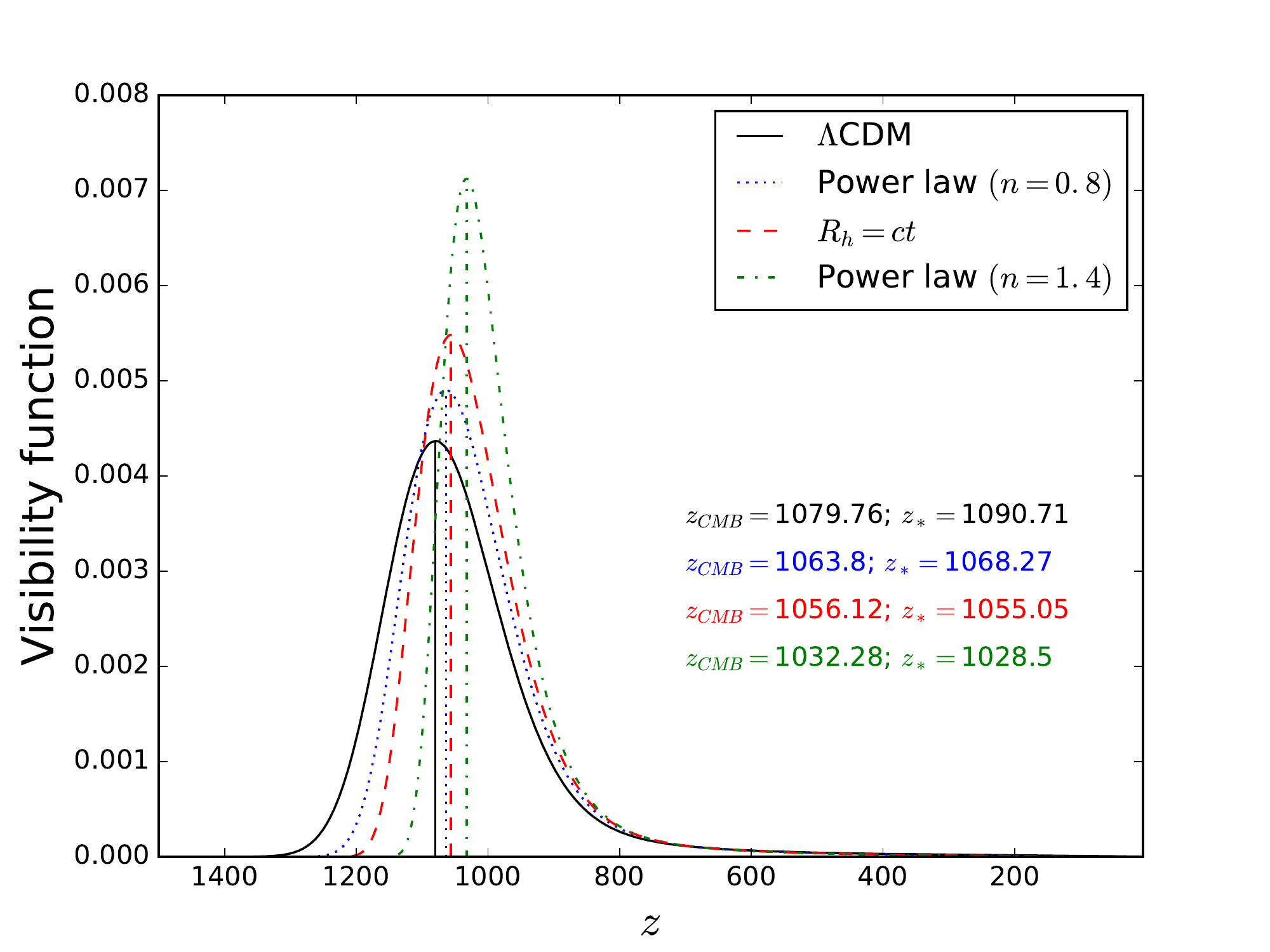}
\caption{Visibility function as a function of the redshift for
  $\Lambda$CDM (black), $R_h=ct$ cosmology (red) and $n=0.8,\,1.4$
  power law cosmologies (blue and green, respectively). We show the redshift
  of the CMB computed with two different definitions (see the text for details).}\label{fig0}
\end{figure}

In Fig.\,\ref{fig0} we show the visibility
functions for $\Lambda$CDM, $R_h=ct$ cosmology and two power law
cosmologies ($n=0.8$ and $n=1.4$). In this case we have fixed the cosmological parameters to $\Lambda$CDM present-day values:
$Y=0.249,\,\Omega_m=0.3089,\,\Omega_b=0.0485976,\,N_{\text{eff}}=3.04$
and $H_0=67.74\,\text{km}\,\text{s}^{-1}\text{Mpc}^{-1}$\,\cite{Planck2015table}. However, we
have checked that any variation of 25\% in one of these parameters has
a negligible impact on the redshift of the CMB (less than 0.6\%) for a
fixed model and that it has no influence in our study. Even if the
redshift of the CMB does not change significantly with the parameters,
it does change with the model; therefore we fix $z_*=1090.71$ for
$\Lambda$CDM and $z_*=1055.05$ for the $R_h=ct$ cosmology. Concerning
the power law cosmology, since the redshift changes significantly with
$n$, we interpolate $z_*$ as a function of $n$.

The redshift of the baryon drag epoch can be
defined in two analogous ways. We first consider the definition given
by a drag visibility function\,\cite{Hu96},

\begin{equation}
g_d(z)=\dot{\tau}_d(z)e^{-\tau_d(z)}\,,
\end{equation}
where the drag optical depth is given by,

\begin{equation}
\tau_d(z)=\int_0^z\frac{\dot{\tau}(z')}{R_b(z')}\,\text{d}z'\,.
\end{equation}

We denote the maximum of this drag visibility function $z_{drag}$. The
second definition (the one adopted by the Planck
Collaboration\,\cite{Planck2015table}) is given by the redshift at
which the drag optical depth equals 1. We denote it $z_d$. As before,
we use $z_d$ to be consistent with Planck, but we keep both
definitions for
illustrative and comparative purposes.

\begin{figure}
\includegraphics[scale=.45]{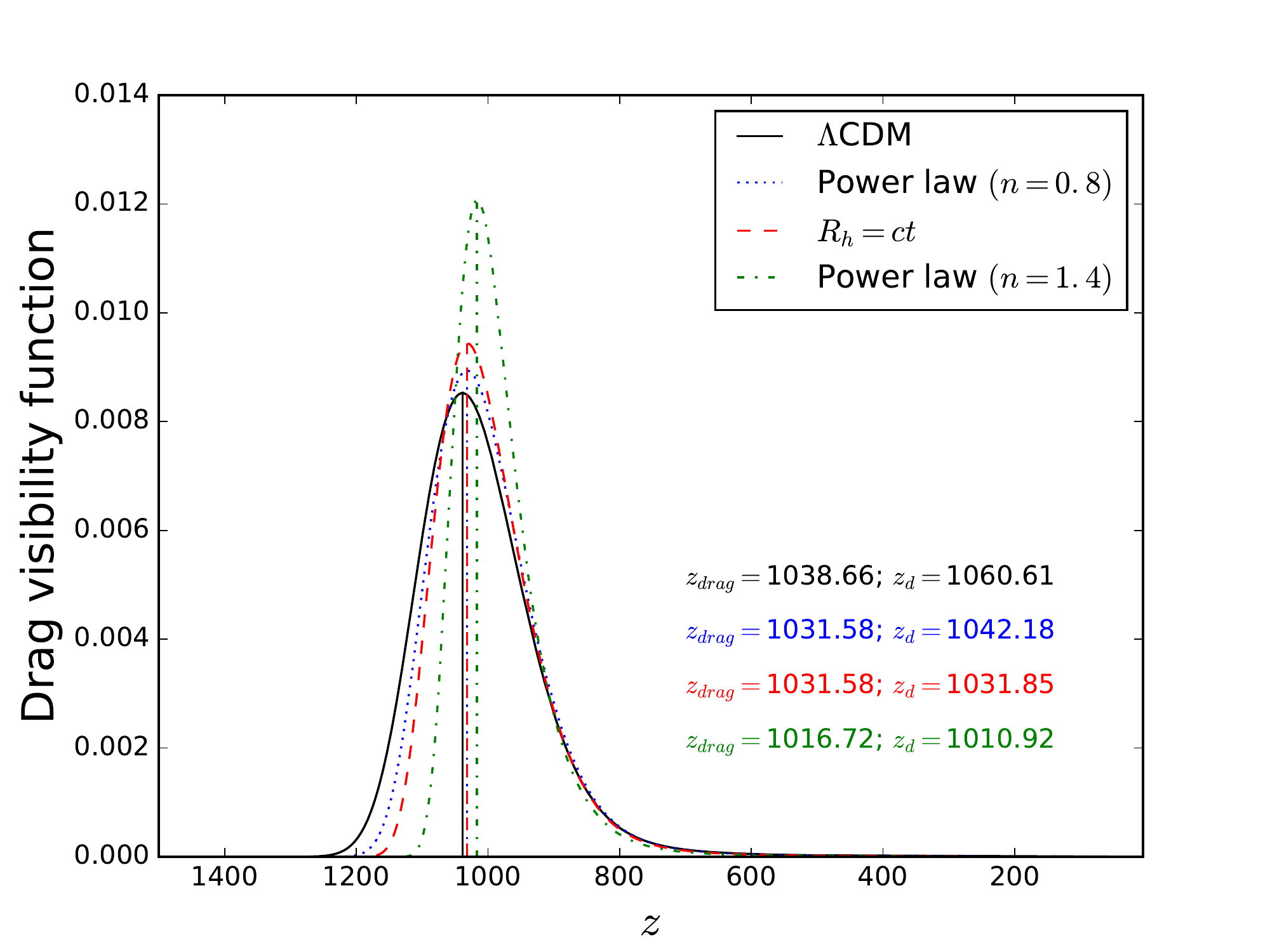}
\caption{Drag visibility function as a function of the redshift for
  $\Lambda$CDM (black), $R_h=ct$ cosmology (red) and $n=0.8,\,1.4$
  power law cosmologies (blue and green, respectively). The redshift
  of the drag epoch computed with two different definitions is presented
  (see the text for details).}\label{fig00}
\end{figure}

In Fig.\,\ref{fig00} we show the drag visibility functions for
the same models that appear in Fig.\,\ref{fig0}. The cosmological
parameters are fixed to the same present-day
values\,\cite{Planck2015table} and we have also checked that any
variation of 25\% in one of the parameters does not lead to
significant changes in our results. Therefore, we fix $z_d=1060.61$
for $\Lambda$CDM and $z_d=1031.85$ for the $R_h=ct$ cosmology. As for
$z_*$ we observe that $z_d$ changes significantly with the exponent of
the power law cosmology;
thus, we interpolate $z_d$ as a function of $n$.

No extra cosmological or nuisance parameters are considered when
including the CMB data.

\section{RESULTS}\label{section6}

In Table \ref{table1} we present the best-fit values obtained for
the different cosmological and nuisance parameters of the models
studied with the different probes used. In Table \ref{table2} we show
the results of the goodness of fit and model comparisons. More
specifically, we report the number of parameters of the model, the
number of data points used, the minimum value for the $\chi^2$
function, the goodness of fit statistic and the exponential of the
differences $\Delta$AICc$/2$ and $\Delta$BIC$/2$.

\begin{figure}
\includegraphics[scale=.45]{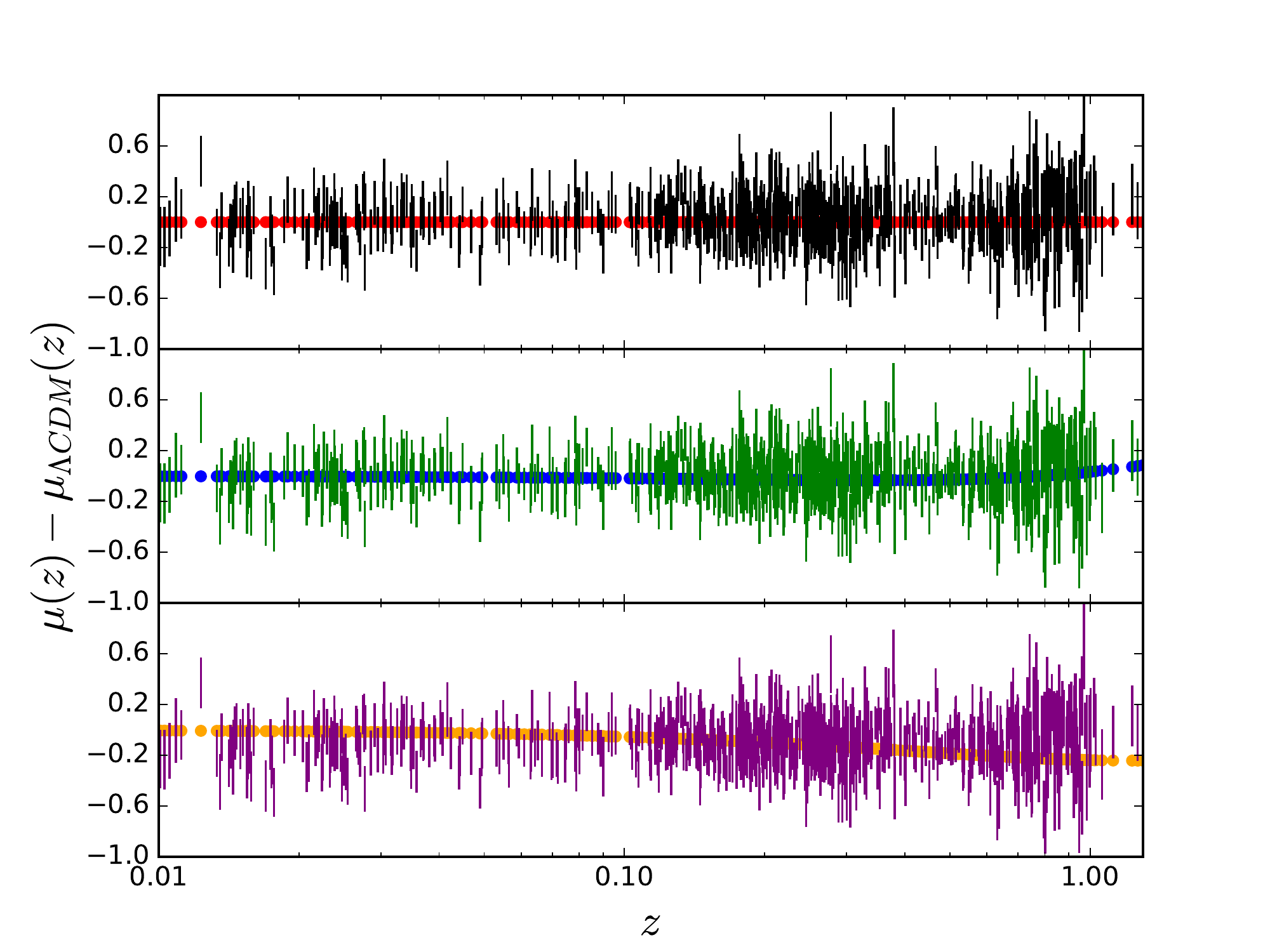}
\caption{Fit from the three models under study to the SNIa data. All
  the plots show the residuals with respect to the prediction from
  $\Lambda$CDM with the best-fit values. Top panel: SNIa measurements
  standardized to $\Lambda$CDM (black) and $\Lambda$CDM prediction
  (red) as a function of the redshift. Central panel: SNIa measurements
  standardized to power law cosmology (green) and power law cosmology
  prediction (blue) as a function of the redshift. Bottom panel: SNIa
  measurements standardized to $R_h=ct$ cosmology (purple) and
  $R_h=ct$ cosmology prediction (orange) as a function of the
  redshift. For each model we marginalize over the nuisance parameters.}\label{fig1}
\end{figure}

Focusing first on the SNIa alone, Fig.\,\ref{fig1} provides the residuals to the best-fit
(normalized to the $\Lambda$CDM model) for each model.
 $\Lambda$CDM provides a very
good fit to the data $(P(\chi^2,\nu)=0.915$), as well as the power law
cosmology $(P(\chi^2,\nu)=0.915$, with $n=1.55\pm 0.13)$. Although the $R_h=ct$ cosmology
provides a slightly worse fit $(P(\chi^2,\nu)=0.644)$, it is still
acceptable. However, it is highly disfavored when considering the
model comparison statistics
$(\text{exp}(\Delta\text{AICc}/2)=1.308\times 10^{-8}$ and
exp$(\Delta\text{BIC}/2)=1.291\times 10^{-7})$. Despite the fact that the $R_h=ct$ model has fewer
parameters than $\Lambda$CDM, the $\chi^2$ difference is large enough
$(\Delta\chi^2=38.33)$ to compensate for the preference of the $R_h=ct$ model coming from the
Occam factor of the AIC and the BIC. By Occam factor we mean here
the non-$\chi^2$ term in Eqs.\,(\ref{aicc}) and\,(\ref{bic}).

\begin{figure}
\includegraphics[scale=.45]{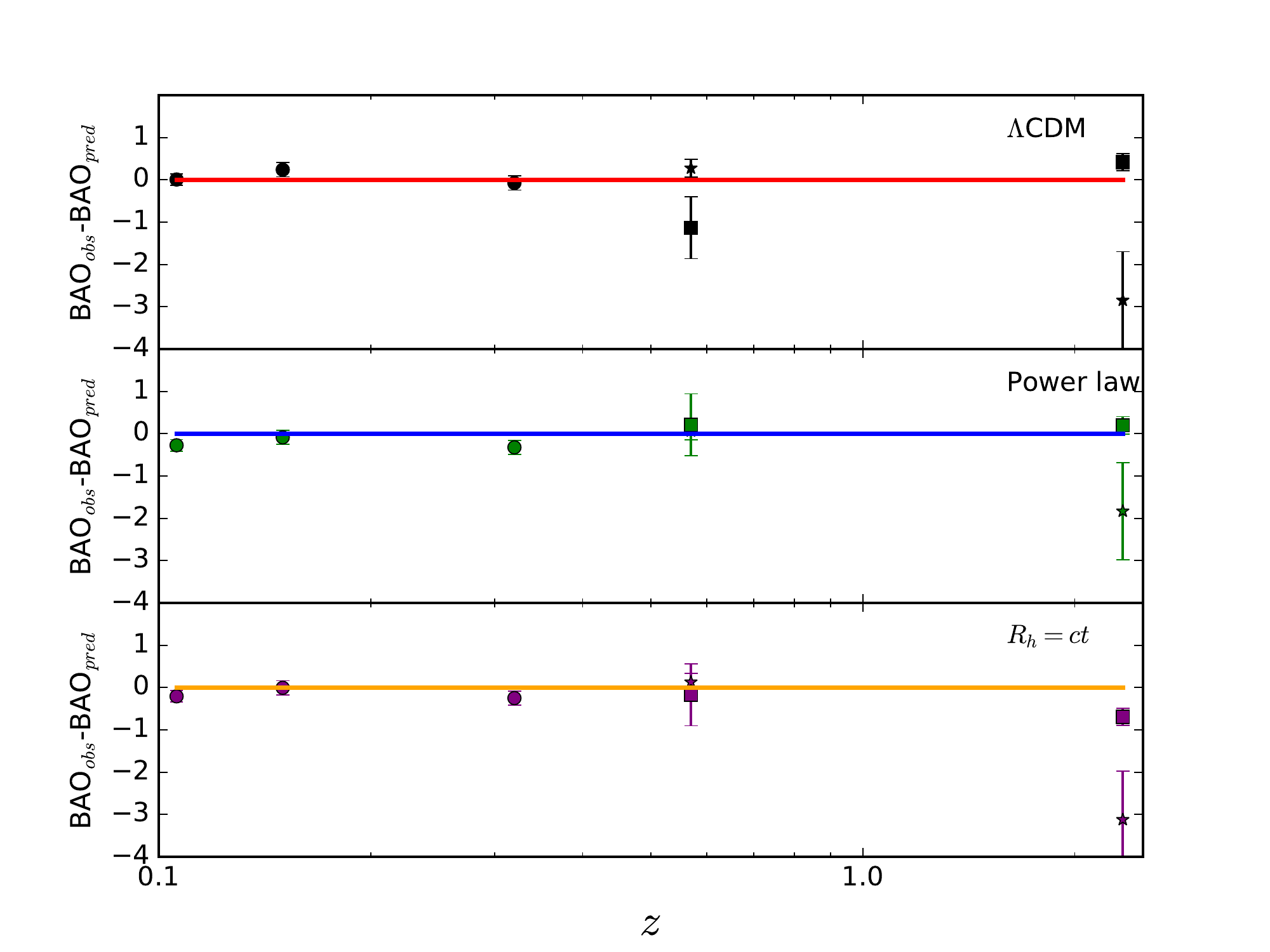}
\caption{Fit from the three models under study to the BAO data.  Each plot shows the residuals with
  respect to the corresponding model. The
  isotropic measurements of the BAO are represented with a circle and
  their observable is $D_V(z)/r_d$,
  while the stars stand for the radial measurements with observable
  $r(z)/r_d$ and the squares
  stand for the transverse measurements with observable $c/(H(z)r_d)$. Top panel: BAO measurements (black)
  and $\Lambda$CDM prediction (red) as a function of the
  redshift. Central panel: BAO measurements (green) and power law
  cosmology prediction (blue) as a function of the redshift. Bottom
  panel: BAO measurements (purple) and $R_h=ct$ cosmology prediction
  (orange) as a function of the redshift. }\label{fig2}
\end{figure}

In Fig.\,\ref{fig2}  we present the residuals of the fit to the BAO data alone from the three
models under study. From the top panel we
immediately see that $\Lambda$CDM is not a good fit to BAO data
$(P(\chi^2,\nu)=0.088)$. This tension has already been
noted in the literature\,\cite{Aubourg,Shafer} and is due to the anisotropic
Lyman-$\alpha$ forest BAO measurements at high redshift $(z=2.34)$. Since SNIa and BAO prefer
similar values of $\Omega_m$, no extra tension appears when
combining these probes. The  power law cosmology provides a better fit to BAO data than $\Lambda$CDM
$(P(\chi^2,\nu)=0.531)$ implying preference of the power law cosmology over $\Lambda$CDM from the model comparison
statistics $(\text{exp}(\Delta\text{AICc}/2)=$exp$(\Delta\text{BIC}/2)=15.198)$. Regarding the $R_h=ct$ model,  the
fit is worse than for $\Lambda$CDM $(P(\chi^2,\nu)=0.016)$, but the
difference of $\chi^2$ with respect to
$\Lambda$CDM is nearly compensated by the Occam factor, so that the
model has commensurate values of the AICc and the BIC: $\text{exp}(\Delta\text{AICc}/2)=0.385$ and
exp$(\Delta\text{BIC}/2)=0.125$, respectively.  

\begin{figure*}
\includegraphics[scale=.45]{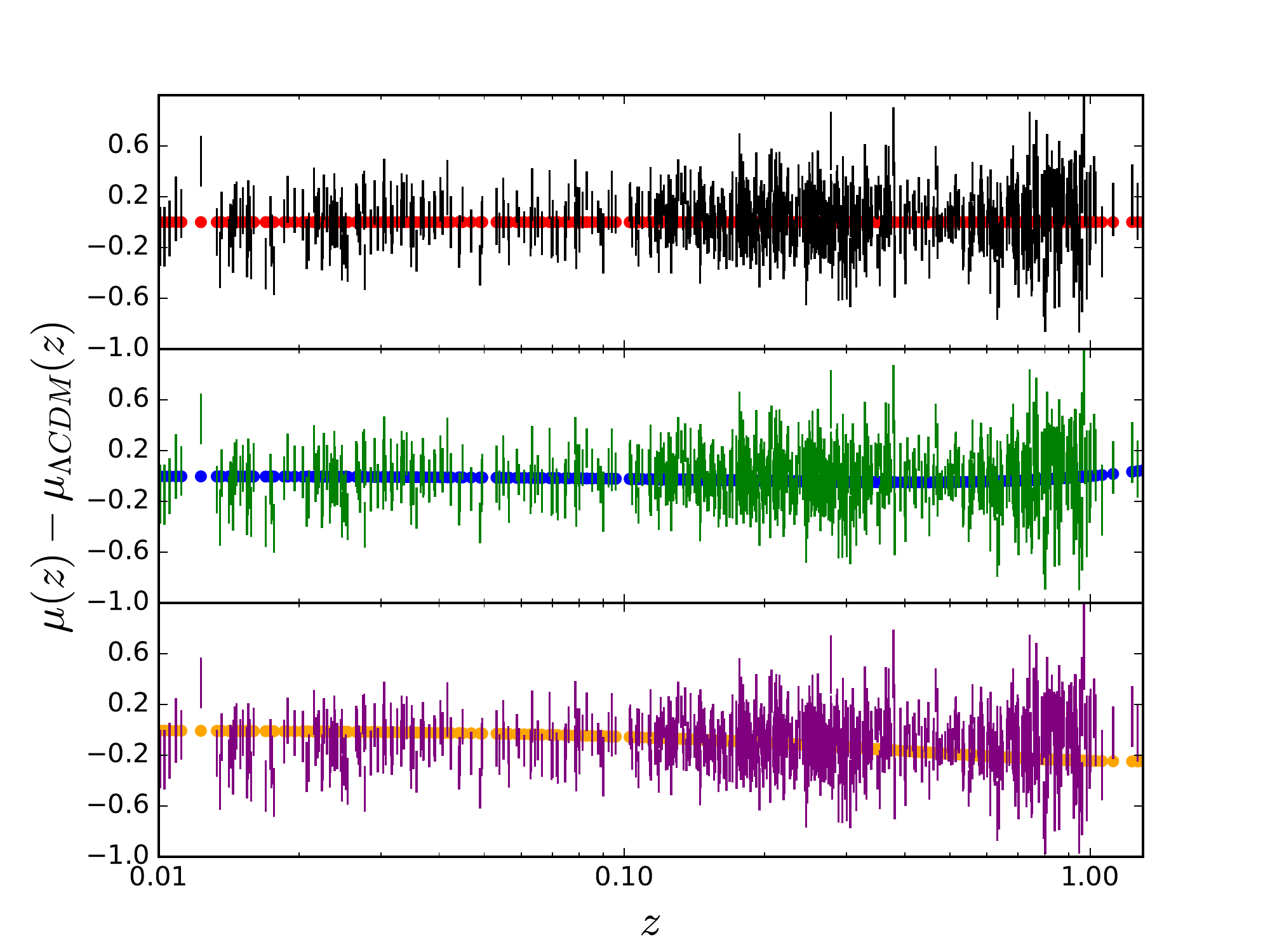}\,\includegraphics[scale=.45]{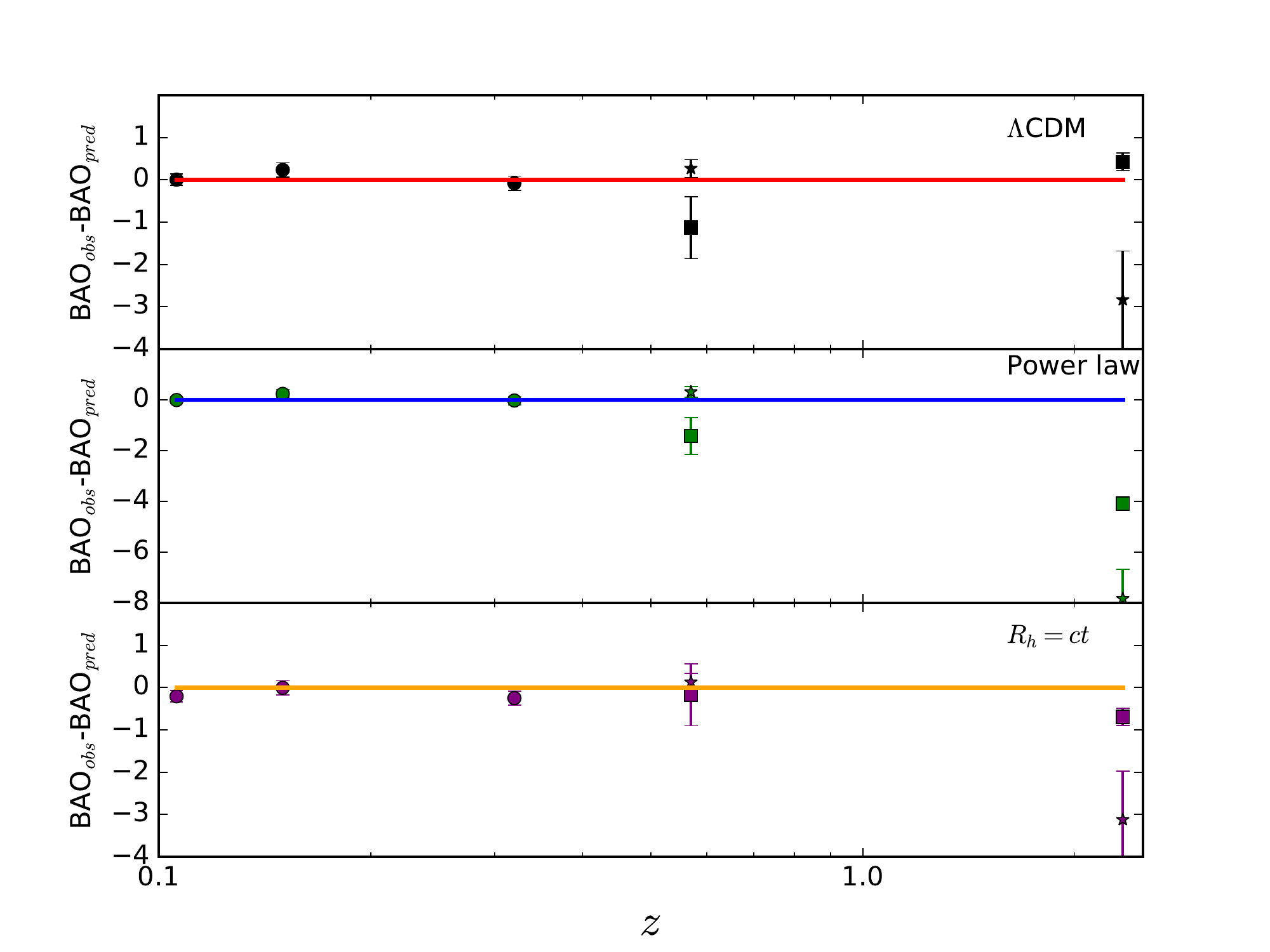}
\caption{Fit from the three models under study to the SNIa and BAO
  data; i.e. the parameter values of the models are the best-fit
  values from SNIa+BAO data. Left plot: SNIa residuals with respect to the prediction from $\Lambda$CDM with the best-fit
  values, for the three models under study (see Fig.\,\ref{fig1}). Right plot: BAO residuals with
  respect to the model under study (see Fig.\,\ref{fig2}).}\label{fig3}
\end{figure*}

In Fig.\,\ref{fig3} we show the results from fitting the three
models to SNIa and BAO data simultaneously. In the left panel we present the fits
from the models to SNIa data using the best-fit values obtained from
both SNIa and BAO data. These results are very
similar to the ones obtained for SNIa alone (Fig.\,\ref{fig1}),
showing that adding the BAO does not affect the SNIa-related
parameters. In the right panel of Fig.\,\ref{fig3} we show the fit from the models to BAO data, using the SNIa+BAO
best-fit values for the parameters. We notice that the power
law cosmology provides a slightly worse fit than when considering BAO
data alone (Fig.\,\ref{fig2}). Looking at the goodness of fit for SNIa
and BAO data, we find that the power law cosmology provides a slightly
worse fit $(P(\chi^2,\nu)=0.833$) than the $\Lambda$CDM
$(P(\chi^2,\nu)=0.898)$, which is also the case for the $R_h=ct$
cosmology $(P(\chi^2,\nu)=0.546)$. Despite the small difference
between the power law cosmology and the $\Lambda$CDM fits, the model
comparison statistics tell us that the latter is preferred $(\text{exp}(\Delta\text{AICc}/2)=$exp$(\Delta\text{BIC}/2)=0.0036)$. The $R_h=ct$ cosmology is even more strongly
disfavored with respect to $\Lambda$CDM than when considering SNIa
data alone $(\text{exp}(\Delta\text{AICc}/2)=6.251\times 10^{-10}$ and
exp$(\Delta\text{BIC}/2)=6.184\times 10^{-9})$. 

From these results (the best-fit values are nearly all consistent with
\cite{Shafer} within 1$\sigma$) we deduce that $R_h=ct$ is very
disfavored with respect to $\Lambda$CDM, while the power law cosmology is
slightly disfavored with respect to $\Lambda$CDM.

\begin{figure}
\includegraphics[scale=.45]{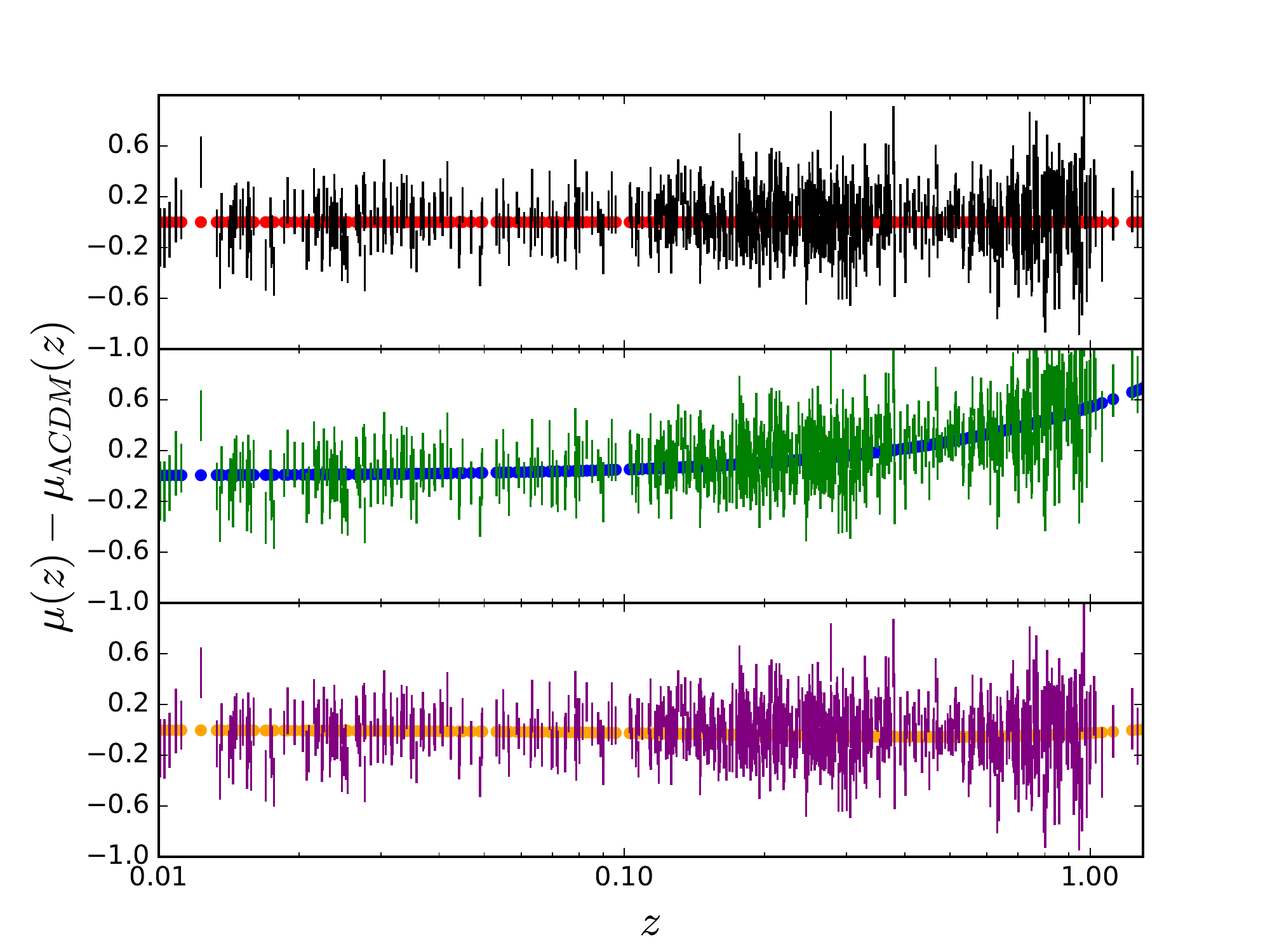}
\caption{Fit from the three models under study to the SNIa data
  allowing for some evolution with the redshift. All the plots show the
  residuals with respect to the prediction from $\Lambda$CDM with the
  best-fit values (see Fig.\,\ref{fig1}). The introduction of
  some evolution with the redshift modifies the observed $\mu(z)$
  giving a good fit for all the models.}\label{fig4}
\end{figure}

In order to be more conservative we allow for some SNIa evolution with the
redshift [Eq.\,(\ref{ev})]. In Fig.\,\ref{fig4} we have the results for
SNIa data. We can observe that now all the models provide a very good
fit to the data. Interestingly, the evolution nuisance parameter is
nearly consistent with 0 for $\Lambda$CDM and the power law cosmology,
while it is clearly non-null for the $R_h=ct$ cosmology. This is
completely consistent since the $\Lambda$CDM and the power law cosmology
were already able to provide a good fit without evolution, while the
$R_h=ct$ needed this nuisance term in order to correctly fit the
data. From the model comparison statistics we can deduce that there is
no clear preference for one model over another. 

\begin{figure*}
\includegraphics[scale=.45]{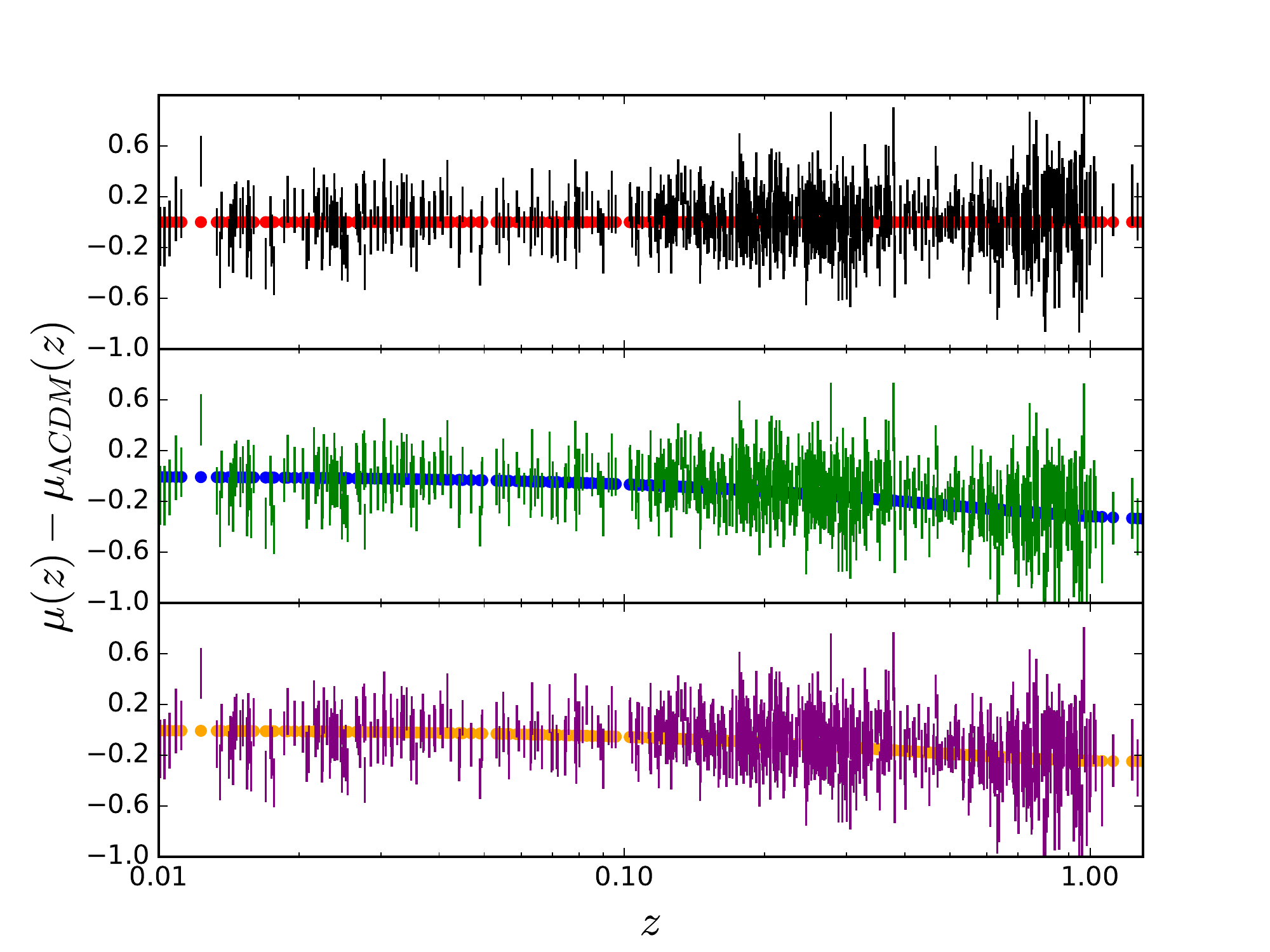}\,\includegraphics[scale=.45]{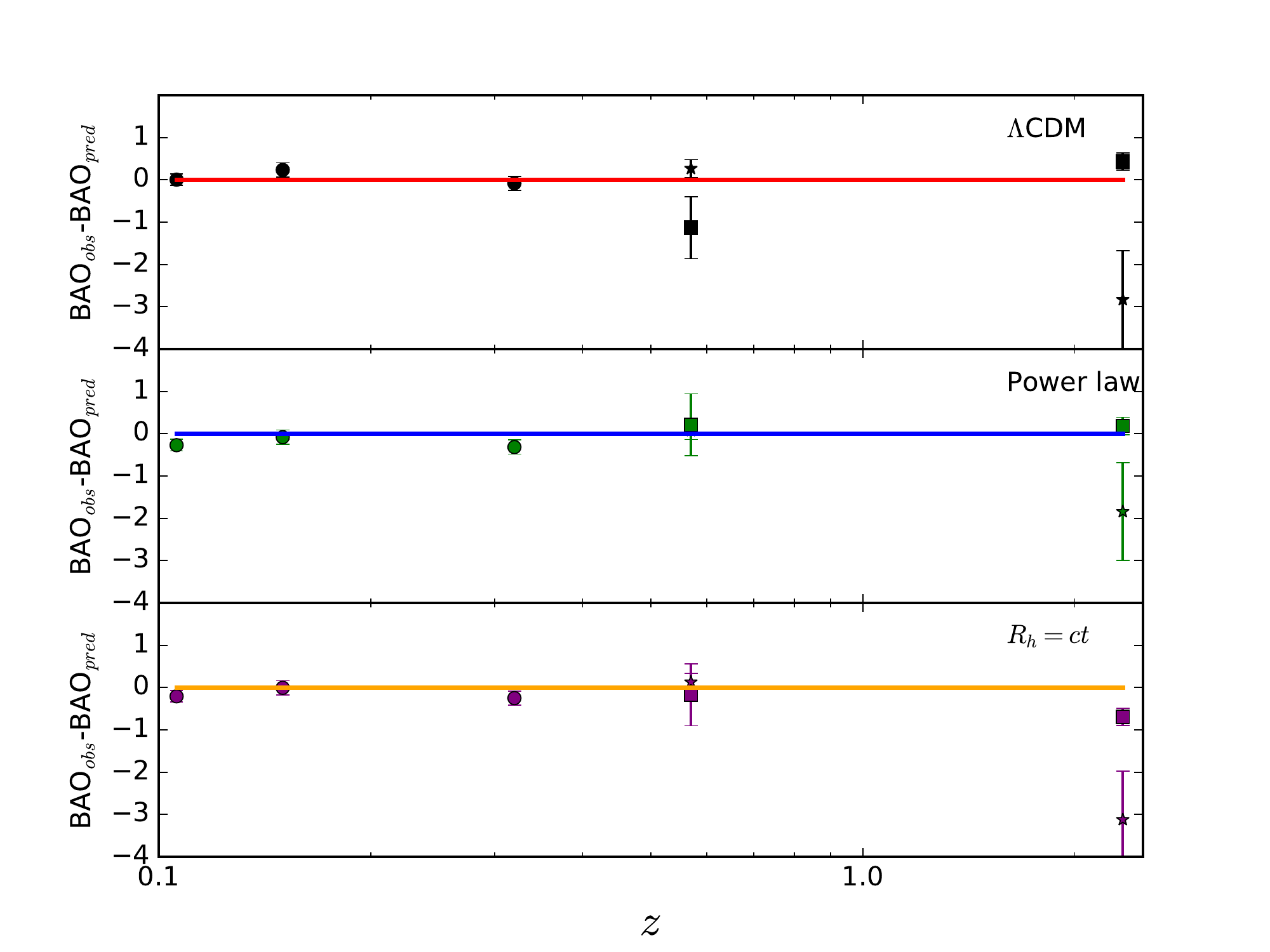}
\caption{Fit from the three models under study to the SNIa and BAO
  data allowing for redshift evolution for the SNIa; i.e. the
  parameter
  values of the models are the best-fit values from SNIa+evolution+BAO
  data. Left plot: SNIa residuals with respect to the
  prediction from $\Lambda$CDM with the best-fit values (see Fig.\,\ref{fig1}). Right plot: BAO residuals with respect to the model under
  study (see Fig.\,\ref{fig2}). Allowing for some redshift
  evolution for SNIa provides a good fit for the three models to both
  SNIa and BAO data.}\label{fig5}
\end{figure*}

Now we can combine the SNIa data (allowing for evolution) with the BAO
data. The results are shown in Fig.\,\ref{fig5}. Contrary to what we
have seen in Fig.\,\ref{fig3}, adding the BAO does modify the SNIa-related parameter values, but we still obtain a very good fit to the
SNIa data using the best-fit values obtained from SNIa+BAO data and
allowing for evolution. Concerning the fit to BAO data, using this
combination of data to determine the best-fit values, we recover the
results obtained with BAO data alone (Fig.\,\ref{fig2}). This shows that when we
relax the redshift independence for SNIa, the power in model selection
from the combination of SNIa and BAO weakens. As the power law
cosmology was slightly preferred over $\Lambda$CDM when considering
BAO data alone, it is not surprising that it is also the case
here $(\text{exp}(\Delta\text{AICc}/2)=$exp$(\Delta\text{BIC}/2)=3.421)$. Concerning the $R_h=ct$ cosmology, the Occam factor is nearly as
important as the $\chi^2$ difference and it leads to only a marginal
preference for the $\Lambda$CDM $(\text{exp}(\Delta\text{AICc}/2)=0.046$ and
exp$(\Delta\text{BIC}/2)=0.455)$.

From these results we can deduce that adding a redshift evolution in
the SNIa as a nuisance parameter leads to no clear preference of one
model over another.

\begin{figure}
\includegraphics[scale=.45]{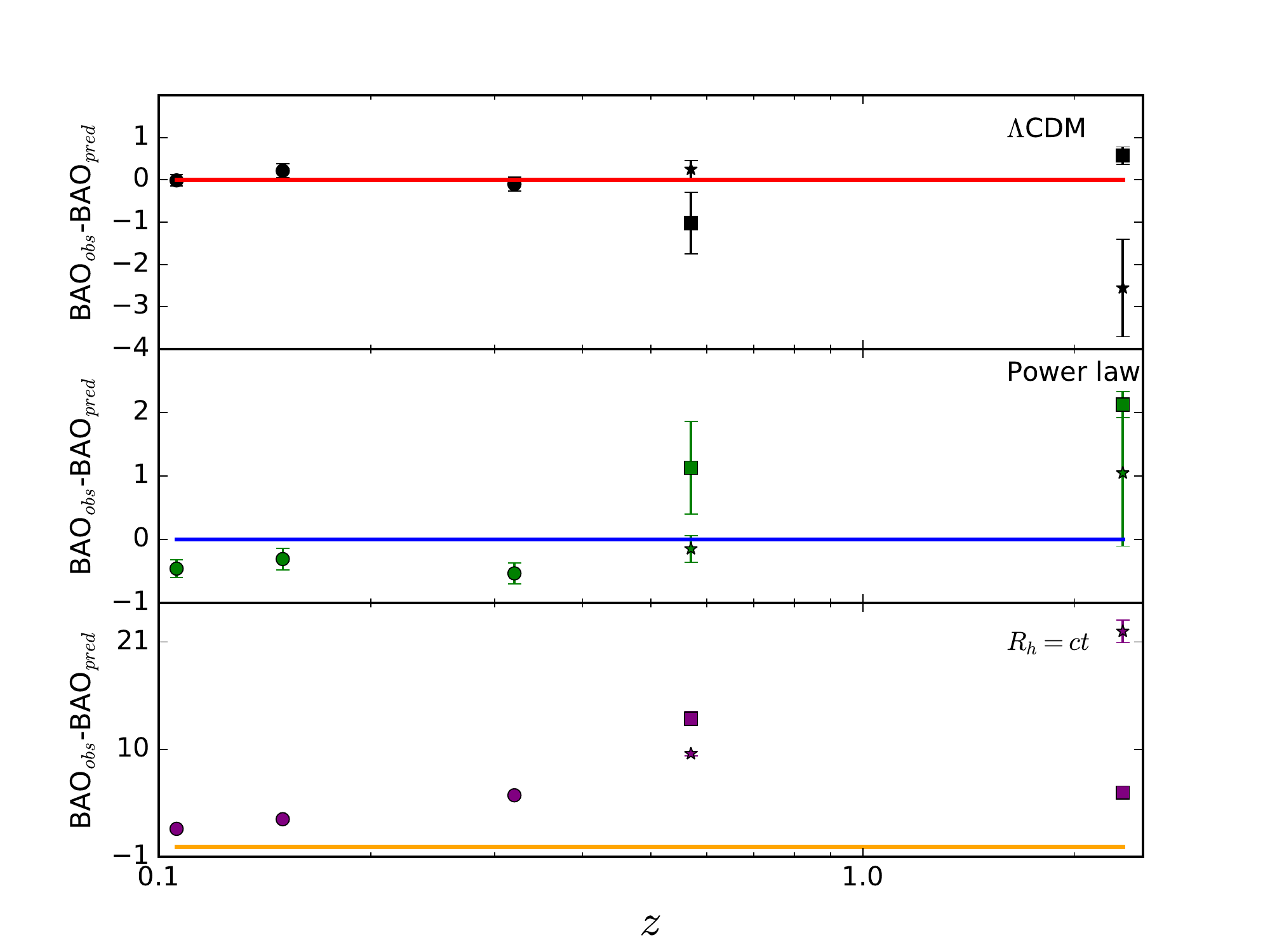}
\caption{Fit from the three models under study to the BAO and CMB
  data. All the plots show the BAO residuals with respect to the model
  under study (see Fig.\,\ref{fig2}). The introduction of the
  CMB data strongly degrades (notice the increase in the Y-axis limits
  and the small size of the error bars) the fit to BAO data for the power law and
the $R_h=ct$ cosmologies.}\label{fig6}
\end{figure}

We finally consider the addition of CMB data. Notice that in
this work we
cannot combine CMB and SNIa data, since we rely on the BAO scale to
introduce the CMB scale (see Sec.\,\ref{CMB}); therefore we always
need to consider BAO data when including the CMB. The results
for BAO and CMB data are shown in Fig.\,\ref{fig6} and in Table\,\ref{table3}. In the plot
we have the results for BAO data with the best-fit values obtained
with BAO and CMB data. In the table we present the value of
$\ell_a$ for each model with the BAO and CMB data best-fit values. We can see that there is no evolution in the $\Lambda$CDM BAO fit
when we add the CMB information to determine the best-fit values, as
expected. However, adding the CMB information is crucial for the power
law and the $R_h=ct$ cosmologies, since the fit to the BAO data is
strongly degraded [$P(\chi^2,\nu)=1.2\times 10^{-3}$ and
$P(\chi^2,\nu)=4.4\times 10^{-14}$, respectively]. From the model
comparison point of view, the power law cosmology is disfavored $(\text{exp}(\Delta\text{AICc}/2)=$exp$(\Delta\text{BIC}/2)=0.0029)$ and
the $R_h=ct$ cosmology is strongly disfavored
$(\text{exp}(\Delta\text{AICc}/2)=1.680\times 10^{-14}$ and
exp$(\Delta\text{BIC}/2)=7.349\times 10^{-15})$ with respect to
$\Lambda$CDM. 

\begin{figure*}
\includegraphics[scale=.45]{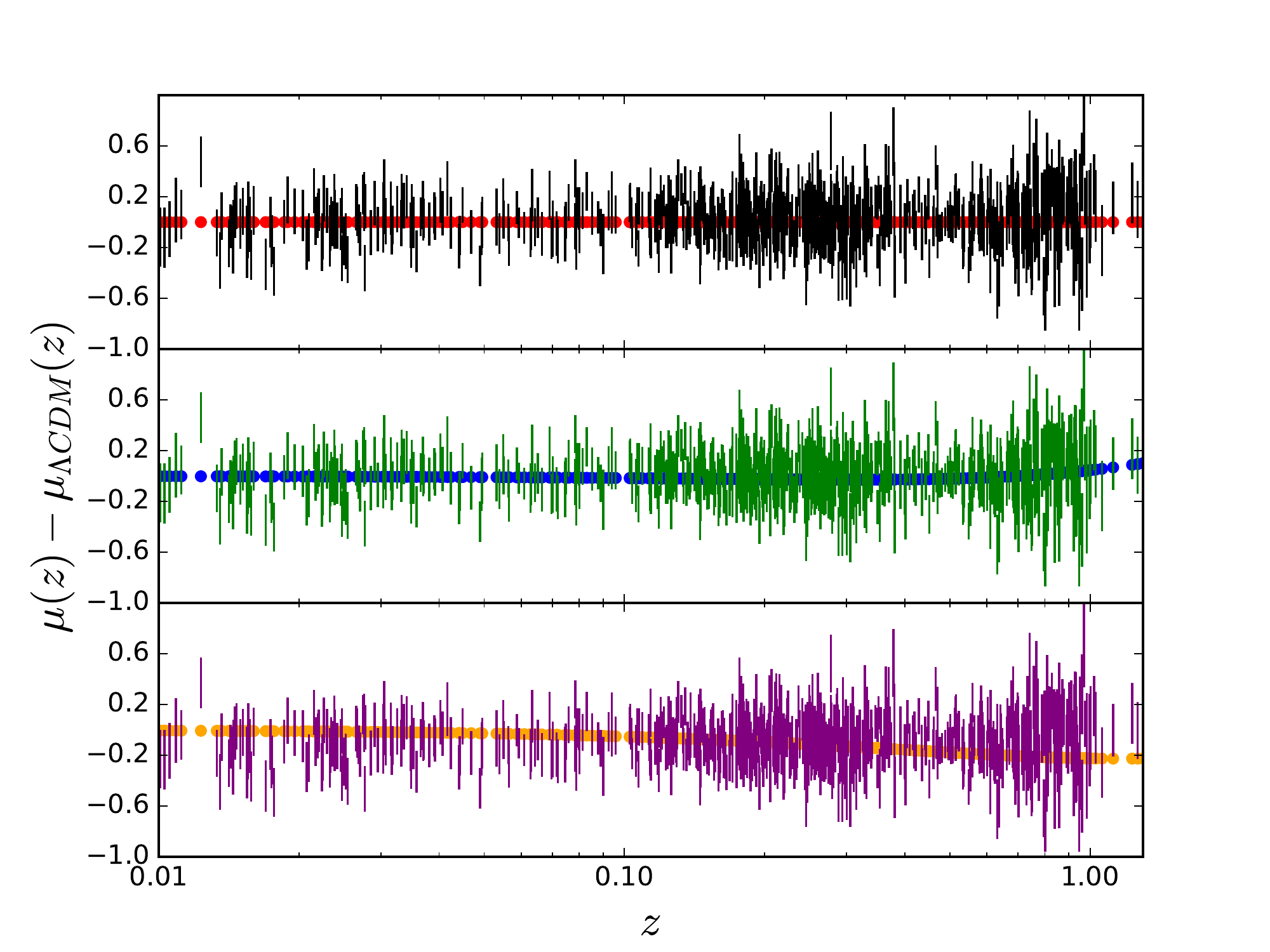}\,\includegraphics[scale=.45]{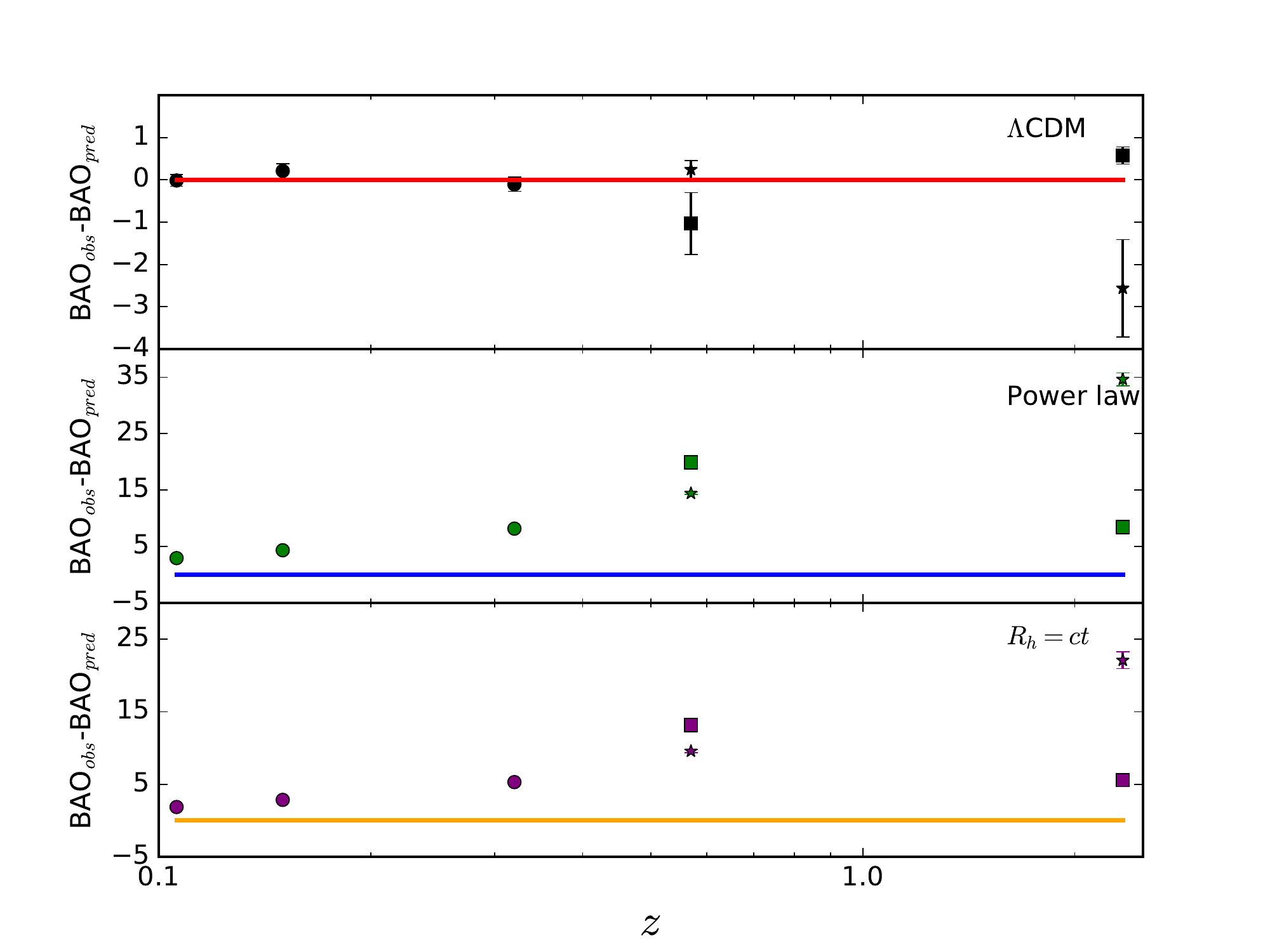}
\caption{Fit from the three models under study to the SNIa, BAO and
  CMB data; i.e. the parameter values of the models are the best-fit
  values from SNIa+BAO+CMB data. Left plot: SNIa residuals with respect to the prediction from $\Lambda$CDM with the
  best-fit values (see Fig.\,\ref{fig1}). Right plot: BAO residuals with
  respect to the model under study (see Fig.\,\ref{fig2}).}\label{fig7}
\end{figure*}

We can now combine the information from the three probes: SNIa, BAO
and CMB. The results are presented in Fig.\,\ref{fig7} and in Table\,\ref{table3}. From the left plot
we can see that adding the CMB information does not affect the fit
to SNIa (see the left panel of Fig.\,\ref{fig5}). However, it completely
degrades the fit to the BAO data for the power law and the $R_h=ct$
cosmologies (see the right panel of Figs.\,\ref{fig5} and\,\ref{fig6}). In terms of model comparison the power law
cosmology is very disfavored $(\text{exp}(\Delta\text{AICc}/2)=$exp$(\Delta\text{BIC}/2)=2.501\times 10^{-15})$ and the $R_h=ct$ cosmology is extremely
disfavored $(\text{exp}(\Delta\text{AICc}/2)=3.598\times 10^{-23}$ and
exp$(\Delta\text{BIC}/2)=3.562\times 10^{-22})$ with respect to
$\Lambda$CDM. It is important to notice here that the $\chi^2$ and the
$P(\chi^2,\nu)$ obtained for the $R_h=ct$ and the power law
cosmologies are acceptable, but the model criteria tell us that these
models are highly improbable. This is due to the introduction of SNIa
data. Both models provide an acceptable fit to these data; so, when
including so many data points, the global fit is still acceptable. However,
the model criteria are essentially sensitive to the exponential of the
difference of $\chi^2$, so they can distinguish between different
models approximately fitting the data. It is a clear example between
the difference of correctly fitting the data and being better than
another model. 

\begin{figure*}
\includegraphics[scale=.45]{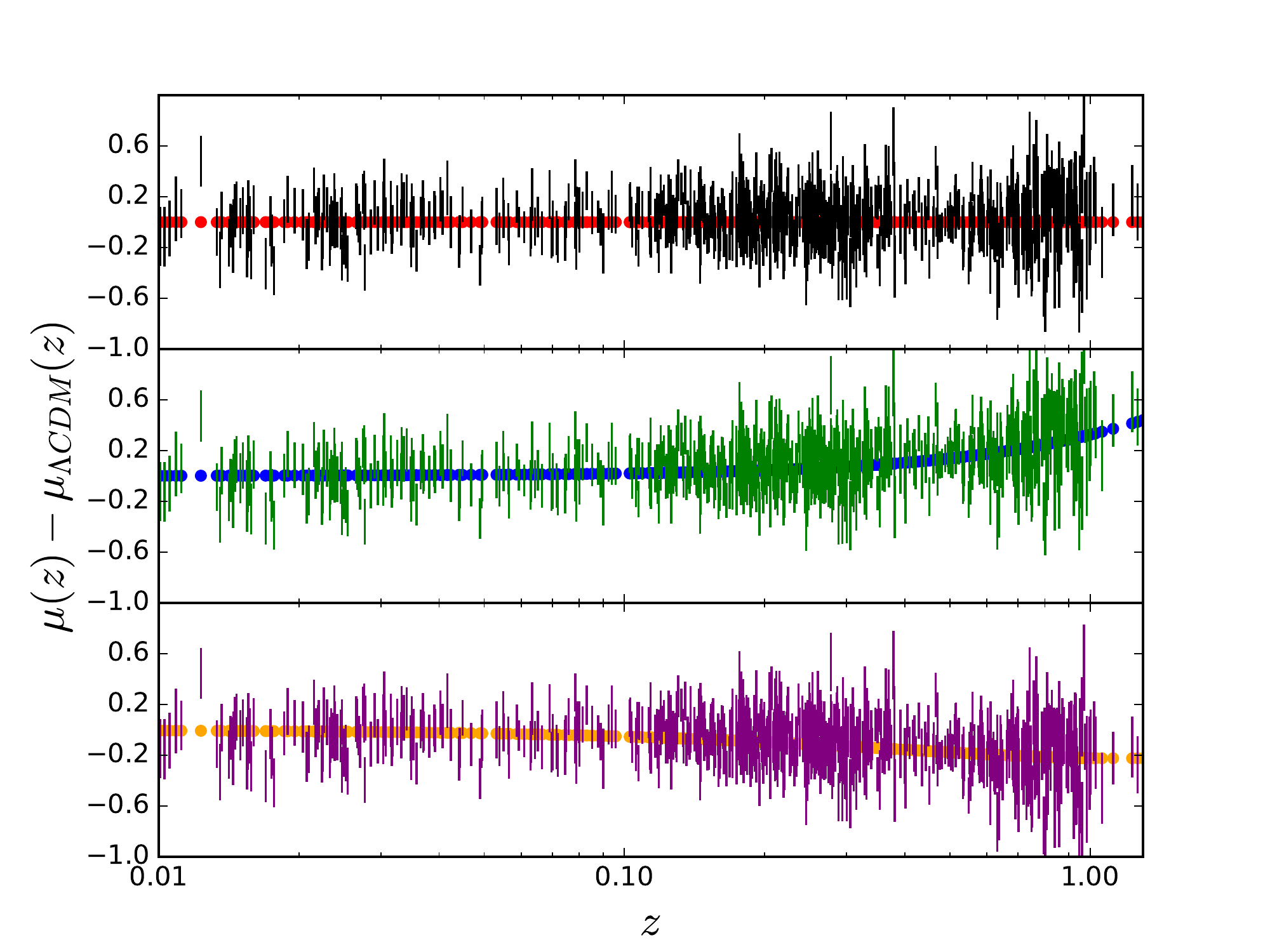}\,\includegraphics[scale=.45]{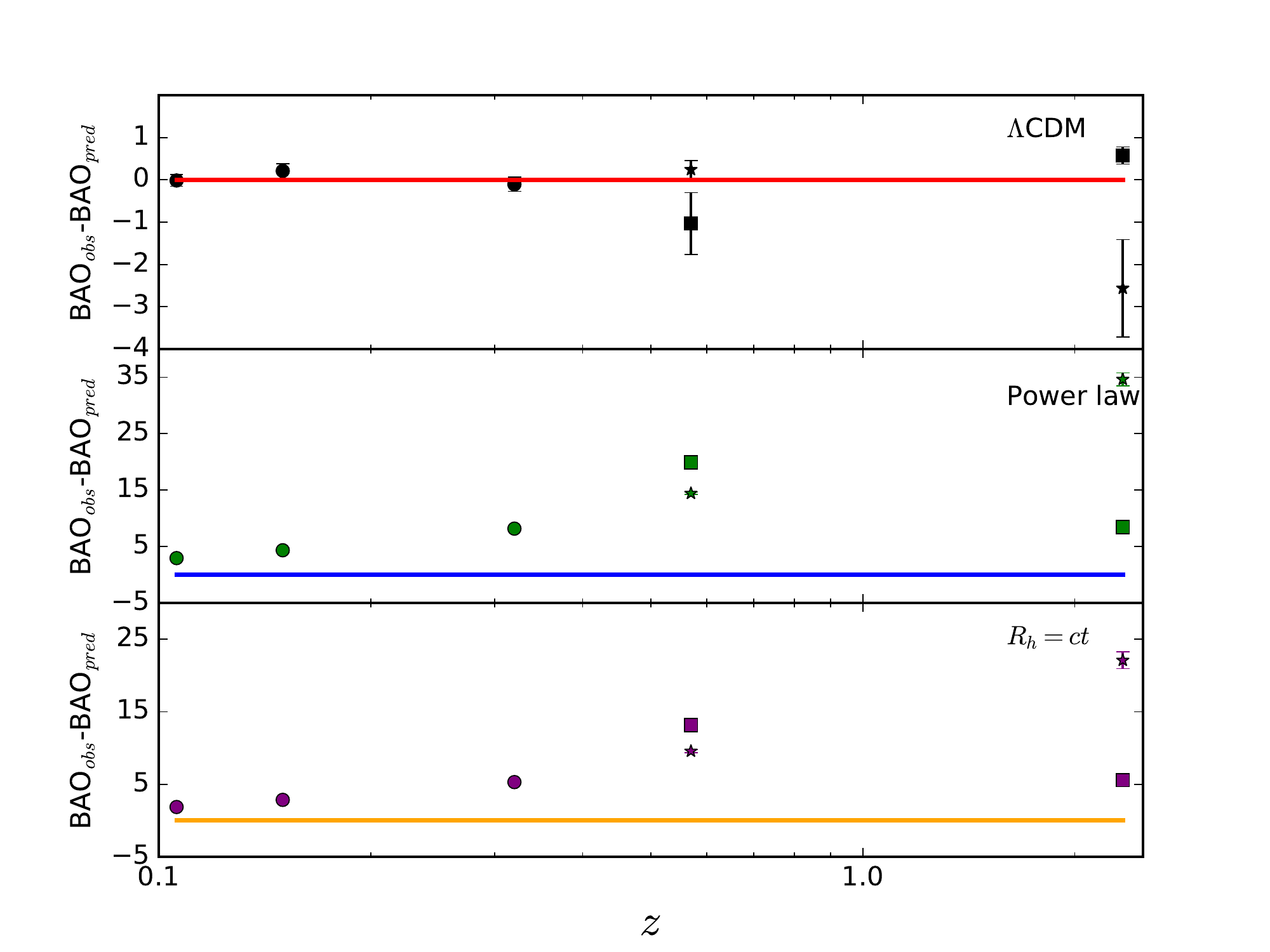}
\caption{Fit from the three models under study to the SNIa, BAO and
  CMB data and allowing for redshift evolution for the SNIa; i.e. the
  parameter values of the models are the best-fit values from
  SNIa+evolution+BAO+CMB data. Left plot: SNIa residuals 
  with respect to the prediction from $\Lambda$CDM with the best-fit
  values (see Fig.\,\ref{fig1}). Right plot: BAO residuals
  with
  respect to the model under study (see 
  Fig.\,\ref{fig2}). Allowing for some redshift evolution for the SNIa
is not sufficient to compensate for the effect of the CMB, and we
remain with a poor fit for the power law and $R_h=ct$ cosmologies.}\label{fig8}
\end{figure*}

For completeness and in order to be as conservative as possible, we
also consider a redshift evolution of SNIa. The results are shown in
Fig.\,\ref{fig8} and in Table\,\ref{table3}. From the left plot we notice that adding the
evolution leads to very good fits to SNIa data. However, from the right plot we can see that the redshift evolution in SNIa is not
sufficient to compensate for the effect of the CMB; thus the power law and
$R_h=ct$ cosmologies are not able to correctly fit the BAO data. The
results from the model comparison still remain clear, showing that
$\Lambda$CDM is very strongly preferred over the power law
$(\text{exp}(\Delta\text{AICc}/2)=$exp$(\Delta\text{BIC}/2)=3.127\times
10^{-15})$ and the $R_h=ct$ $(\text{exp}(\Delta\text{AICc}/2)=2.363\times 10^{-15}$ and
exp$(\Delta\text{BIC}/2)=2.332\times 10^{-14})$ cosmologies.

\begin{sidewaystable}[t]
\caption{Best-fit values for the cosmological and nuisance parameters
  of the studied models with different cosmological probes.}\label{table1}
\begin{center}
\resizebox{\textwidth}{!}{
\begin{tabular}{cc|c|c|c|c|c|c|c|c|}
\cline{3-10}
& & $\Omega_m$ & $n$ & $r_d\times H_0/c$ & $\alpha$ & $\beta$ & $M$ &
                                                                      $\Delta
                                                                      M$
  & $\epsilon$\\
\cline{1-10}
\multicolumn{1}{|c}{\multirow{8}{*}{$\Lambda$CDM}} &
                                                     \multicolumn{1}{|l|}{SNIa}
  & $0.295\pm 0.034$ & - & - & $0.1412\pm 0.0066$ & $3.101\pm 0.081$ & $24.110\pm 0.023$ & $-0.070\pm 0.023$ & - \\

\multicolumn{1}{|c}{} & \multicolumn{1}{|l|}{BAO} & $0.289\pm 0.021$ &
                                                                       - & $0.03377\pm 0.00057$ & - & - & - & - & - \\

\multicolumn{1}{|c}{} & \multicolumn{1}{|l|}{SNIa+BAO} & $0.291\pm 0.018$ & - & $0.03373\pm 0.00050$ & $0.1413\pm 0.0066$ & $3.103\pm 0.080$ & $24.110\pm 0.019$ & $-0.070\pm 0.023$ & - \\

\multicolumn{1}{|c}{} & \multicolumn{1}{|l|}{SNIa+ev} & $0.49\pm 0.17$ & - & - & $0.1410\pm 0.0066$ & $3.100\pm 0.081$ & $24.120\pm 0.025$ & $-0.070\pm 0.023$ & $-0.23\pm 0.19$ \\

\multicolumn{1}{|c}{} & \multicolumn{1}{|l|}{SNIa+ev+BAO} & $0.291\pm 0.021$ & - & $0.03373\pm 0.00057$ & $0.1413\pm 0.0066$ & $3.103\pm 0.081$ & $24.110\pm 0.022$ & $-0.070\pm 0.023$ & $-0.001\pm 0.055$ \\

\multicolumn{1}{|c}{} & \multicolumn{1}{|l|}{BAO+CMB} & $0.307\pm 0.011$ & - & $0.03348\pm 0.00049$ & - & - & - & - & - \\

\multicolumn{1}{|c}{} & \multicolumn{1}{|l|}{SNIa+BAO+CMB} & $0.306\pm 0.010$ & - & $0.03353\pm 0.00045$ & $0.1411\pm 0.0066$ &$3.098\pm 0.080$ & $24.110\pm 0.018$ & $-0.070\pm 0.023$ & - \\

\multicolumn{1}{|c}{} & \multicolumn{1}{|l|}{SNIa+ev+BAO+CMB} & $0.308\pm 0.011$ & - & $0.03345\pm 0.00049$ & $0.1413\pm 0.0066$ & $3.103\pm 0.081$ & $24.110\pm 0.022$ & $-0.070\pm 0.023$ & $-0.024\pm 0.049$ \\

\hline

\multicolumn{1}{|c}{\multirow{8}{*}{Power law}} &
                                                     \multicolumn{1}{|l|}{SNIa}
  & - & $1.55\pm 0.13$ & - & $0.1408\pm 0.0066$ & $3.100\pm 0.081$ & $24.130\pm 0.023$ & $-0.071\pm 0.023$ & - \\

\multicolumn{1}{|c}{} & \multicolumn{1}{|l|}{BAO} & - &
                                                                       $0.908\pm 0.019$ & $0.02963\pm 0.00040$ & - & - & - & - & - \\

\multicolumn{1}{|c}{} & \multicolumn{1}{|l|}{SNIa+BAO} & - & $1.46\pm 0.11$ & $0.03313\pm 0.00059$ & $0.1404\pm 0.0066$ & $3.094\pm 0.080$ & $24.140\pm 0.023$ & $-0.072\pm 0.023$ & - \\

\multicolumn{1}{|c}{} & \multicolumn{1}{|l|}{SNIa+ev} & - & $3.7\pm 5.1$ & - & $0.1410\pm 0.0066$ & $3.100\pm 0.081$ & $24.120\pm 0.024$ & $-0.071\pm 0.023$ & $0.31\pm 0.32$ \\

\multicolumn{1}{|c}{} & \multicolumn{1}{|l|}{SNIa+ev+BAO} & - & $0.910\pm 0.019$ & $0.02967\pm 0.00040$ & $0.1404\pm 0.0066$ & $3.100\pm 0.081$ & $24.150\pm 0.022$ & $-0.072\pm 0.023$ & $-0.354\pm 0.049$ \\

\multicolumn{1}{|c}{} & \multicolumn{1}{|l|}{BAO+CMB} & - & $0.7345\pm 0.0025$ & $0.02758\pm 0.00029$ & - & - & - & - & - \\

\multicolumn{1}{|c}{} & \multicolumn{1}{|l|}{SNIa+BAO+CMB} & - & $1.545\pm 0.067$ & $0.402\pm 0.054$ & $0.1408\pm 0.0066$ &$3.100\pm 0.080$ & $24.130\pm 0.019$ & $-0.071\pm 0.023$ & - \\

\multicolumn{1}{|c}{} & \multicolumn{1}{|l|}{SNIa+ev+BAO+CMB} & - & $3.330\pm 0.089$ & $1.019\pm 0.068$ & $0.1409\pm 0.0066$ & $3.100\pm 0.081$ & $24.120\pm 0.022$ & $-0.071\pm 0.023$ & $0.292\pm 0.047$ \\

\hline

\multicolumn{1}{|c}{\multirow{8}{*}{$R_h=ct$}} &
                                                     \multicolumn{1}{|l|}{SNIa}
  & - & - & - & $0.1382\pm 0.0066$ & $3.073\pm 0.080$ & $24.230\pm 0.017$ & $-0.077\pm 0.023$ & - \\

\multicolumn{1}{|c}{} & \multicolumn{1}{|l|}{BAO} & - &
                                                                       - & $0.03045\pm 0.00031$ & - & - & - & - & - \\

\multicolumn{1}{|c}{} & \multicolumn{1}{|l|}{SNIa+BAO} & - & - & $0.03045\pm 0.00031$ & $0.1382\pm 0.0066$ & $3.073\pm 0.080$ & $24.230\pm 0.017$ & $-0.077\pm 0.023$ & - \\

\multicolumn{1}{|c}{} & \multicolumn{1}{|l|}{SNIa+ev} & - & - & - & $0.1405\pm 0.0066$ & $3.100\pm 0.081$ & $24.140\pm 0.022$ & $-0.072\pm 0.023$ & $-0.277\pm 0.046$ \\

\multicolumn{1}{|c}{} & \multicolumn{1}{|l|}{SNIa+ev+BAO} & - & - & $0.03045\pm 0.00031$ & $0.1405\pm 0.0066$ & $3.100\pm 0.081$ & $24.140\pm 0.022$ & $-0.072\pm 0.023$ & $-0.277\pm 0.046$ \\

\multicolumn{1}{|c}{} & \multicolumn{1}{|l|}{BAO+CMB} & - & - & $0.083770\pm 0.000036$ & - & - & - & - & - \\

\multicolumn{1}{|c}{} & \multicolumn{1}{|l|}{SNIa+BAO+CMB} & - & - & $0.083770\pm 0.000036$ & $0.1382\pm 0.0066$ &$3.073\pm 0.080$ & $24.230\pm 0.017$ & $-0.077\pm 0.023$ & - \\

\multicolumn{1}{|c}{} & \multicolumn{1}{|l|}{SNIa+ev+BAO+CMB} & - & - & $0.083770\pm 0.000036$ & $0.1405\pm 0.0066$ & $3.100\pm 0.081$ & $24.140\pm 0.022$ & $-0.072\pm 0.023$ & $-0.277\pm 0.046$ \\

\hline

\end{tabular}
}
\end{center}
\end{sidewaystable}

\begin{table*}[t]
\caption{Goodness of fit and model comparison between the models
  studied with the different cosmological probes considered. The last
  two columns for $\Lambda$CDM and power law cosmology are combined
  because exp$(\Delta\text{AICc}/2)=\text{exp}(\Delta\text{BIC}/2)$ in
  these cases (see the text in Sec.\,\ref{modelcomparison}).}\label{table2}
\begin{center}
\resizebox{\textwidth}{!}{
\begin{tabular}{cc|c|c|c|c|c|c|}
\cline{3-8}
& & $k$ & $N$ & $\chi^2_{min}$ & $P(\chi^2,\nu)$ & exp$(\Delta \text{AICc}/2)$
  & exp$(\Delta \text{BIC}/2)$\\

\cline{1-8}
\multicolumn{1}{|c}{\multirow{8}{*}{$\Lambda$CDM}} &
                                                     \multicolumn{1}{|l|}{SNIa}
  & 5 & 740 & 682.89 & 0.915 & \multicolumn{2}{c|}{1} \\

\multicolumn{1}{|c}{} & \multicolumn{1}{|l|}{BAO} & 2 & 7 & 9.57 & 0.088 & \multicolumn{2}{c|}{1} \\

\multicolumn{1}{|c}{} & \multicolumn{1}{|l|}{SNIa+BAO} & 6 & 747 & 692.49 & 0.898 & \multicolumn{2}{c|}{1} \\

\multicolumn{1}{|c}{} & \multicolumn{1}{|l|}{SNIa+ev} & 6 & 740 & 681.90 & 0.916 & \multicolumn{2}{c|}{1} \\

\multicolumn{1}{|c}{} & \multicolumn{1}{|l|}{SNIa+ev+BAO} & 7 & 747 & 692.48 & 0.893 & \multicolumn{2}{c|}{1} \\

\multicolumn{1}{|c}{} & \multicolumn{1}{|l|}{BAO+CMB} & 2 & 8 & 10.36 & 0.110 & \multicolumn{2}{c|}{1} \\

\multicolumn{1}{|c}{} & \multicolumn{1}{|l|}{SNIa+BAO+CMB} & 6 & 748 & 693.36 & 0.899 & \multicolumn{2}{c|}{1} \\

\multicolumn{1}{|c}{} & \multicolumn{1}{|l|}{SNIa+ev+BAO+CMB} & 7 & 748 & 693.13 & 0.895 & \multicolumn{2}{c|}{1} \\

\hline

\multicolumn{1}{|c}{\multirow{8}{*}{Power law}} &
                                                     \multicolumn{1}{|l|}{SNIa}
  & 5 & 740 & 682.90 & 0.915 & \multicolumn{2}{c|}{0.998} \\

\multicolumn{1}{|c}{} & \multicolumn{1}{|l|}{BAO} & 2 & 7 & 4.13 & 0.531 & \multicolumn{2}{c|}{15.198} \\

\multicolumn{1}{|c}{} & \multicolumn{1}{|l|}{SNIa+BAO} & 6 & 747 & 703.71 & 0.833 & \multicolumn{2}{c|}{0.0036} \\

\multicolumn{1}{|c}{} & \multicolumn{1}{|l|}{SNIa+ev} & 6 & 740 & 682.20 & 0.914 & \multicolumn{2}{c|}{0.860} \\

\multicolumn{1}{|c}{} & \multicolumn{1}{|l|}{SNIa+ev+BAO} & 7 & 747 & 690.03 & 0.905 & \multicolumn{2}{c|}{3.421} \\

\multicolumn{1}{|c}{} & \multicolumn{1}{|l|}{BAO+CMB} & 2 & 8 & 22.07 & 0.0012 & \multicolumn{2}{c|}{0.0029} \\

\multicolumn{1}{|c}{} & \multicolumn{1}{|l|}{SNIa+BAO+CMB} & 6 & 748 &
                                                                       760.61
                               & 0.310 & \multicolumn{2}{c|}{2.501$\times 10^{-15}$} \\

\multicolumn{1}{|c}{} & \multicolumn{1}{|l|}{SNIa+ev+BAO+CMB} & 7 &
                                                                    748
              & 759.93 & 0.307 & \multicolumn{2}{c|}{3.127$\times 10^{-15}$}\\

\hline

\multicolumn{1}{|c}{\multirow{8}{*}{$R_h=ct$}} &
                                                     \multicolumn{1}{|l|}{SNIa}
  & 4 & 740 & 721.22 & 0.644 & 1.308$\times 10^{-8}$ & 1.291$\times 10^{-7}$ \\

\multicolumn{1}{|c}{} & \multicolumn{1}{|l|}{BAO} & 1 & 7 & 15.68 & 0.016 & 0.385 & 0.125 \\

\multicolumn{1}{|c}{} & \multicolumn{1}{|l|}{SNIa+BAO} & 5 & 747 &
                                                                   736.90
                               & 0.546 & 6.251$\times 10^{-10}$ &
                                                                  6.184$\times 10^{-9}$ \\

\multicolumn{1}{|c}{} & \multicolumn{1}{|l|}{SNIa+ev} & 5 & 740 & 685.00 & 0.906 & 0.588 & 5.793 \\

\multicolumn{1}{|c}{} & \multicolumn{1}{|l|}{SNIa+ev+BAO} & 6 & 747 & 700.67 & 0.853 & 0.046 & 0.455 \\

\multicolumn{1}{|c}{} & \multicolumn{1}{|l|}{BAO+CMB} & 1 & 8 & 77.53
                               & 4.39$\times 10^{-14}$ & 1.680$\times
                                                         10^{-14}$ &
                                                                     7.349$\times 10^{-15}$ \\

\multicolumn{1}{|c}{} & \multicolumn{1}{|l|}{SNIa+BAO+CMB} & 5 & 748 &
                                                                       798.75
                               & 0.077 & 3.598$\times 10^{-23}$ &
                                                                  3.562$\times 10^{-22}$ \\

\multicolumn{1}{|c}{} & \multicolumn{1}{|l|}{SNIa+ev+BAO+CMB} & 6 &
                                                                    748
              & 762.52 & 0.293 & 2.363$\times 10^{-15}$ & 2.332$\times
  10^{-14}$\\

\hline

\end{tabular}
}
\end{center}
\end{table*}

\begin{table*}[t]
\caption{Value of $\ell_a$ for the different models under study with
  the best-fit values coming from the different combinations of data sets used. The Planck 2015 value
  has been added for comparison.}\label{table3}
\begin{center}
\resizebox{\textwidth}{!}{
\begin{tabular}{c|c|c|c|c|}
\cline{2-5}
 & $\ell_a$ Planck 2015 & $\ell_a$ $\Lambda$CDM & $\ell_a$ Power law
  & $\ell_a$ $R_h=ct$ \\

\cline{1-5}

\multicolumn{1}{|l|}{BAO+CMB} & $301.63\pm 0.15$ & 301.651 & 301.677& 301.649\\
\hline
\multicolumn{1}{|l|}{SNIa+BAO+CMB} & $301.63\pm 0.15$ & 301.591 & 301.856 & 301.649\\
\hline
\multicolumn{1}{|l|}{SNIa+ev+BAO+CMB} & $301.63\pm 0.15$ & 301.529 &
                                                                     301.415
  & 301.649  \\
  
\hline

\end{tabular}
}
\end{center}
\end{table*}

\section{CONCLUSIONS}\label{section7}

In this work we have studied the ability of three different models,
the $\Lambda$CDM, power law cosmology and $R_h=ct$ cosmology, to fit
cosmological data and we have compared these models using two
different model comparison statistics: the Akaike information
criterion and the Bayesian information criterion. We have seen that
all three models are able to fit the data if we only consider SNIa data,
but $R_h=ct$ is disfavored with respect to the $\Lambda$CDM and the power
law cosmology, from a model comparison point of view. Considering BAO
data alone we have observed that the $\Lambda$CDM is
not a good fit to data, due to the anisotropic measurement of the
Lyman-$\alpha$ forest, and we have seen that the power law cosmology is slightly
preferred over $\Lambda$CDM (and significantly preferred over the
$R_h=ct$ cosmology). However, when combining SNIa and BAO
data, the $\Lambda$CDM is preferred over the other models. We have then
considered a possible redshift evolution in SNIa. This has led to an 
excellent fit to SNIa for all the models and, even when adding the BAO
data, there is no clearly preferred model. We have finally considered
the scale information coming from the CMB. In order to use this
information we have made one assumption: the physics driving the
comoving sound horizon at the early Universe in the $R_h=ct$ and power
law cosmologies is the same as in the $\Lambda$CDM model. This
assumption is justified by the existence of radiation and baryon components in the power law and $R_h=ct$
cosmologies, which should lead to an early universe photon-baryon plasma
similar to the one predicted by $\Lambda$CDM. When adding
the scale information from the CMB to BAO and SNIa data we have observed that the goodness of fit
remains the same for $\Lambda$CDM, but it is completely degraded for
the other models. Even adding some evolution to SNIa we have seen that
it is not
sufficient to compensate for the effect of the CMB. This degradation shows
the tension present in the power law and $R_h=ct$ cosmologies between
the BAO scale and the CMB scale, coming from the first peak of the
temperature angular power spectrum. We can conclude that the $\Lambda$CDM
is statistically very strongly preferred over power
law and $R_h=ct$ cosmologies.

\section*{ACKNOWLEDGEMENTS}
This work has been partially supported by the COsmology BEyond SIX
parameters (COBESIX) group in the Origines Constituants \& EVolution
de l’Univers Excellence Laboratory (LabEx OCEVU, Grant
No. ANR-11-LABX-0060) and the A*MIDEX project (Grant No. ANR-11-IDEX-0001-02), funded by the Investissements d’Avenir French government program managed by the Agence Nationale de la Recherche.

\end{document}